\documentclass{JHEP}
\usepackage{epsfig}
\usepackage[active]{srcltx}

\catcode`\@=11
\makeatletter \ifcase\@ptsize \font\teneufm=eufm10
\font\seveneufm=eufm7 \font\fiveeufm=eufm5
\font\teneusm=eusm10 \font\seveneusm=eusm7
\font\fiveeusm=eusm5 \or \font\teneufm=eufm10 scaled
\magstephalf \font\seveneufm=eufm7 \font\fiveeufm=eufm5
\font\teneusm=eusm10 scaled \magstephalf
\font\seveneusm=eusm7 \font\fiveeusm=eusm5 \or
\font\teneufm=eufm10 scaled \magstep1 \font\seveneufm=eufm7
\font\fiveeufm=eufm5 \font\teneusm=eusm10 scaled \magstep1
\font\seveneusm=eusm7 \font\fiveeusm=eusm5 \fi

\newfam\eufmfam \newfam\eusmfam \textfont\eufmfam=\teneufm
\scriptfont\eufmfam=\seveneufm
\scriptscriptfont\eufmfam=\fiveeufm
\textfont\eusmfam=\teneusm \scriptfont\eusmfam=\seveneusm
\scriptscriptfont\eusmfam=\fiveeusm

\def\frak{\ifmmode\let\next\frak@\else
\def\next{\errmessage{Use \string\frak\space only in math
mode}}\fi\next} \def\frak@#1{{\frak@@{#1}}}
\def\frak@@#1{\fam\eufmfam#1} 
\def\sh{\ifmmode\let\next\sh@\else
\def\next{\errmessage{Use \string\sh\space only in math
mode}}\fi\next} \def\sh@#1{{\sh@@{#1}}}
\def\sh@@#1{\fam\eusmfam#1}

\ifcase\@ptsize \font\tenmsa=msam10 \font\sevenmsa=msam7
\font\fivemsa=msam5 \font\tenmsb=msbm10
\font\sevenmsb=msbm7 \font\fivemsb=msbm5 \or
\font\tenmsa=msam10 scaled \magstephalf
\font\sevenmsa=msam7 \font\fivemsa=msam5
\font\tenmsb=msbm10 scaled \magstephalf
\font\sevenmsb=msbm7 \font\fivemsb=msbm5 \or
\font\tenmsa=msam10 scaled \magstep1 \font\sevenmsa=msam7
\font\fivemsa=msam5 \font\tenmsb=msbm10 scaled \magstep1
\font\sevenmsb=msbm7 \font\fivemsb=msbm5 \fi

\newfam\msafam \newfam\msbfam \textfont\msafam=\tenmsa
\scriptfont\msafam=\sevenmsa
\scriptscriptfont\msafam=\fivemsa \textfont\msbfam=\tenmsb
\scriptfont\msbfam=\sevenmsb
\scriptscriptfont\msbfam=\fivemsb

\def\Bbb{\ifmmode\let\next\Bbb@\else
\def\next{\errmessage{Use \string\Bbb\space only in math
mode}}\fi\next} \def\Bbb@#1{{\Bbb@@{#1}}}
\def\Bbb@@#1{\fam\msbfam#1} \def\hexnumber@#1{\ifnum#1<10
\number#1\else \ifnum#1=10 A\else\ifnum#1=11
B\else\ifnum#1=12 C\else \ifnum#1=13 D\else\ifnum#1=14
E\else\ifnum#1=15 F\fi\fi\fi\fi\fi\fi\fi}
\def\msa@{\hexnumber@\msafam} \def\msb@{\hexnumber@\msbfam}
\mathchardef\square="0\msa@03

\makeatother
\newcommand{\beq}{\begin{equation}}
\newcommand{\eeq}{\end{equation}}

\newcommand{\ba}{\begin{array}}
\newcommand{\ea}{\end{array}}
\newcommand{\bea}{\begin{eqnarray}}
\newcommand{\eea}{\end{eqnarray}}
\newcommand{\bean}{\begin{eqnarray*}}
\newcommand{\eean}{\end{eqnarray*}}

\newcommand{\be}{\begin{equation}}
\newcommand{\ee}{\end{equation}}

\newtheorem{theorem}{Theorem}[section]

\newtheorem{remark}[theorem]{Remark}

\newtheorem{proof}{Proof.}

\makeatother

 \def\be{\beta}

\def\be{\begin{equation}}
\def\ee{\end{equation}}
\def\l{\label}

\preprint{MIT-CTP-3224\\ \\ {\tt
hep-th/yymmddd}}

\title{The Splitting of Branes on Orientifold Planes}

\author{Gaetano Bertoldi, Bo Feng, Amihay Hanany,
\footnote{ Research supported in part by the CTP and the LNS of
MIT and the U.S. Department of Energy under cooperative research
agreement \# DE-FC02-94ER40818. A. H. is also supported by an A.
P. Sloan Foundation Fellowship, a DOE OJI award and by the Reed
Fund Award. G. B. is also supported in part by the INFN ``Bruno
Rossi'' Fellowship.}
\\
Center for Theoretical Physics,
\\ Massachusetts Institute of Technology\\ Cambridge MA 02139\\
\email{bertoldi, fengb, hanany@ctp.mit.edu}
}

\abstract{Continuing the study in
\href{http://xxx.lanl.gov/abs/hep-th/0004092}{hep-th/0004092}, we
investigate a non-trivial string dynamical process related to
orientifold planes, i.e., the splitting of physical NS-branes and
$D(p+2)$-branes on orientifold $Op$-planes. Creation or
annihilation of physical $Dp$-branes usually accompanies the
splitting process. In the particular case $p=4$, we use
Seiberg-Witten curves as an independent method to check the
results.}

\begin{document}
\section{Introduction}
The brane setup \cite{Han-Wit} (for a review, see \cite{Giv}) has
been proved to be a very useful tool to relate string theory to
field theory. On one hand, we can use known results in string
theory to get non-trivial new results in field theory. For
instance, mirror symmetry in three dimensions \cite{Int} can be
easily studied by brane setup \cite{Han-Wit,Boer1,
Boer2,Kap,Han2,Mir}. On the other hand, field theory
considerations can be used to predict non-trivial dynamics in
string theory. One of the most known examples is the generation or
annihilation of a $D3$-brane when a $D5$-brane crosses an NS-brane
\cite{Han-Wit}. This prediction and its generalizations were
demonstrated later from various points of view
\cite{Bachas,9705084,9705130,9706142,9708137,9710218,
9711117,0006117}.

\par
In this note, we want to study one prediction about string
dynamics which was found in \cite{Mir} using consistency arguments
within field theory. In \cite{Mir}, in order to construct mirror
pairs by $O3$-planes, two of us made a conjecture on the splitting
of a physical brane (NS-brane and $D5$-brane) on the $O3$-planes.
The general picture of splitting is given in Figure
\ref{fig:picture}. Initially, a 1/2NS-brane with non-vanishing
$X^{789}$ is separated from its image (see part (a)). Then, we
move them towards the orientifold plane at $X^{789}=0$. When they
touch the orientifold, they are on top of each other. Since now
each piece is mirror to itself, they can move along the
$X^6$-direction independently (see part (b)). This process is
called ``splitting''.

\EPSFIGURE[ht]{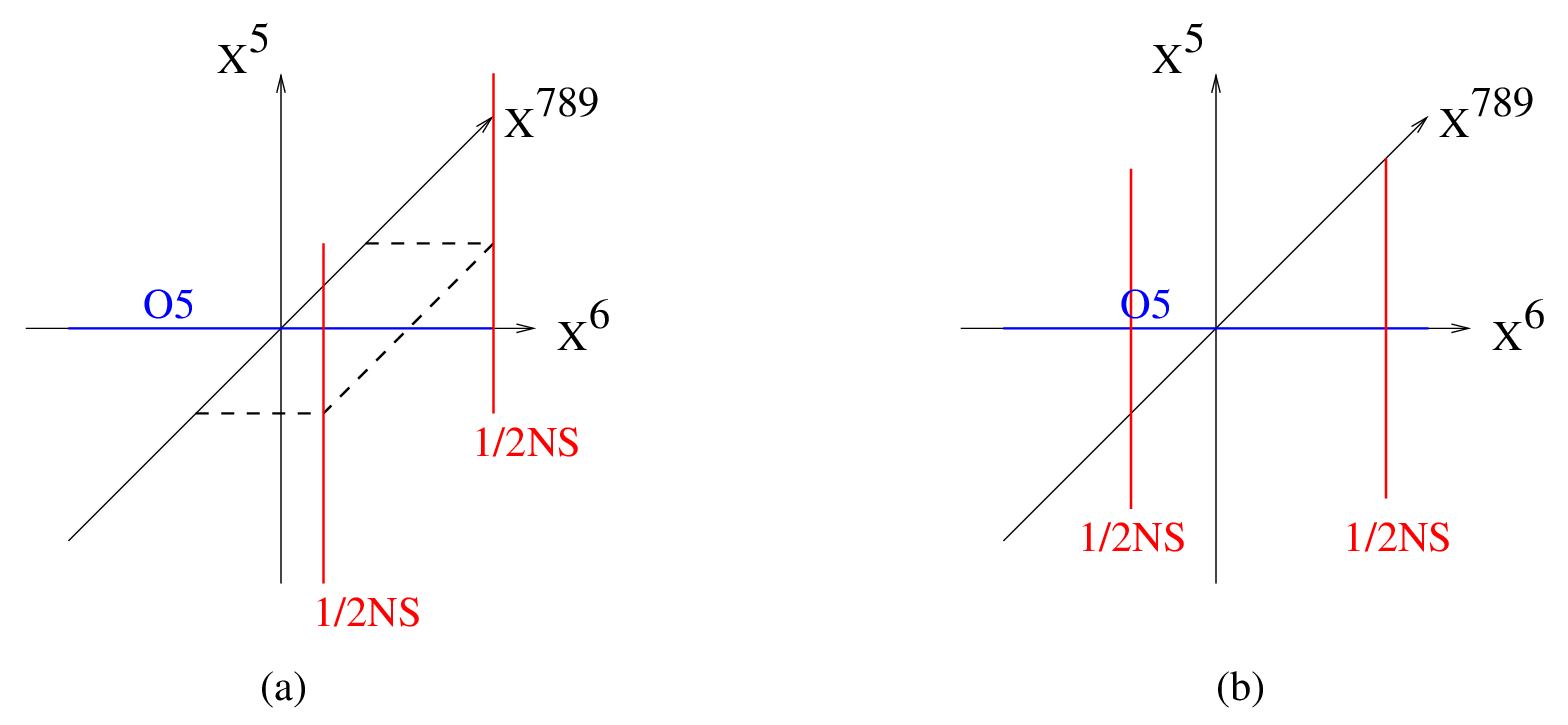,width=14cm}{ (a) 1/2NS-brane and its image
have the same $X^6$ coordinate. (b) After splitting, they have
different $X^6$ coordinates. \label{fig:picture}}

The results of splitting in \cite{Mir} \footnote{ For simplicity,
we will only state the results of the splitting of physical branes
which have no net $D3$-branes ending on them. In order to obtain
the more general case, we can use the same trick given in
\cite{Mir}.} are as follows. A physical $D5$-brane splits without
the generation of branes  on the $O3^+, O3^-,\widetilde{O3^+}$
planes, while a physical $D3$-brane is generated on the
$\widetilde{O3^-}$-plane. The splitting of a physical NS-brane can
be derived from the $D5$-brane case by S-duality. The NS-brane
will split freely, namely without the generation of branes, on
$\widetilde{O3^-}, O3^-,\widetilde{O3^+}$ planes, while a physical
$D3$-brane is generated on the $O3^+$ plane.

We got the above results indirectly by using field theory
arguments. It would be very interesting to gather more evidence to
support them, or even derive them directly. Furthermore, it would
be very nice to discuss how NS-branes and $D(p+2)$-branes split on
$Op$-planes for arbitrary $p$. It turns out that the splitting of
$D(p+2)$-branes on $Op$-planes is universal for $p \leq 6$, namely
the conclusions we have derived for the case of $D5$-branes on
O3-planes hold for all $p$. However, the splitting of NS-branes on
$Op$-planes is more complex and there is no universal behavior.
Each case has to be discussed separately. The results can be
summarized in the following table:

\be
\label{result1}
\begin{tabular}{|c|c|c|c|c|c|c|}  \hline
Type           & $p=1$ & $p=2$ & $p=3$ & $p=4$ & $p=5$ & $p=6$ \\ \hline
$Op^+$         & $0$   & $0$  & $1$ & $ 1$ & $\leq 2$ & $M-4$  \\  \hline
$\widetilde{Op^+}$ & $0$ & $0$ & $0$ & $0$ & $\leq 1$ & $M-7/2$ \\  \hline
$Op^-$  &       $0$ & $0$ & $0$ & $-1$ & $\leq -2$ & $M+4$ \\ \hline
$\widetilde{Op^-}$ & $0$ & $0$ & $0$ & $-1$ & $\leq -2 $ & $M+7/2$ \\ \hline
\end{tabular}
\ee

In the table above, $1$ means that a $Dp$-brane is generated while
$-1$ means that a $Dp$-brane is annihilated. For the annihilation
to be possible, we need enough $Dp$-branes on top of the
orientifold before the splitting. The $M$ for the $p=6$ case is
the cosmological constant which we will discuss in more detail
later. For $p=5$ case, the $\leq 2$ means that any number of
branes which is less than or equal to two is allowed (we use the
convention that a negative number means annihilation).

The above conclusions are derived by several consistency arguments.
In the special case of $p=4$, we can check the result by studying
the Coulomb branch of ${\cal N}=2$ symplectic and orthogonal gauge theories
realized in this brane setup. In particular, we will analyze
the corresponding Seiberg-Witten curves.

\par
The plan of the paper is as follows. In section \ref{two}, we
derive the general splitting rules for both $D$-branes and
NS-branes. In section \ref{three}, we explain how to check these
rules in the specific case of the NS-brane splitting on
$O4$-planes by Seiberg-Witten curves. Then, we introduce the
curves for ${\cal N}=2$ theories with a single symplectic or
orthogonal gauge group. In section \ref{four}, we write down the
curves for a product of symplectic and orthogonal gauge groups and
then study the NS-brane splitting.

\section{The general picture of the splitting}\label{two}
In this section, we will give the general picture of the splitting
of physical $D$-branes and NS-branes on $Op$-planes. We will argue
that the splitting of $D$-branes on orientifolds is universal,
while the splitting of NS-branes is not.

The plan for this section is as follows. In section \ref{twone},
we study the universal splitting of $D$-branes. In section
\ref{2.2}, we observe that for NS-branes, the splitting is not
universal so that we need to discuss each case separately. In
section \ref{minimum}, we use the minimum energy argument to
derive the splitting of NS-branes on $Op$-planes with $p\leq 4$.
In section \ref{O6} and \ref{O5}, we use proper consistency
conditions, i.e., the cosmological constant argument for $p=6$ and
charge conservation for $p=5$, to reach the desired results.

\subsection{The universal splitting of $D$-branes}
\label{twone}

First, let us discuss the splitting of a physical $D(p+2)$-brane
on $Op^\pm, \widetilde{Op^\pm}$-planes for $p \leq 6$
    \footnote{Recently the authors of \cite{Ima} discussed the
    existence of
    $\widetilde{Op^-}$ for $p=6,7,8$. They showed that it is indeed
        possible
    for $p=6$, in the presence of a non-zero cosmological constant,
    but it is not allowed for $p=7,8$. If $\widetilde{O6^-}$ exists,
        there should be $\widetilde{O6^+}$ as well, since crossing
    a 1/2NS-brane changes minus charged planes to plus charged ones
        \cite{EJS,Wit1}.}.
The  $D(p+2)$-brane is extended along the $X^{012 \ldots (p-1)}$
and $X^{789}$ directions, while the $Op$-plane is extended along
the $X^{012 \ldots (p-1)}$ and $X^6$ directions \footnote{We focus
on the case where the brane setup preserves 8 supercharges. }.
There are actually more than four kinds of $O1$-planes, but we
will not discuss all of them in this paper and refer the reader to
\cite{0102095,0103183,Han1}. We will give several arguments to
show that the splitting of a physical $D(p+2)$-brane on
$Op$-planes is universal: the $D(p+2)$-brane will split freely on
$Op^+, Op^-,\widetilde{Op^+}$ planes, while a physical $Dp$-brane
is created on the $\widetilde{Op^-}$-plane \footnote{Further
observations supporting such a universal behaviour will be
presented when we discuss the splitting of NS-branes.}.

\par
The first argument, which was employed in \cite{Mir}, involves the
Higgs mechanism. Let us consider the familiar brane setup given
by two 1/2NS-branes, $k$ physical $Dp$-branes and $n$ physical
$D(p+2)$-branes which preserves 8 supercharges. The $Dp$-branes
are parallel to the $Op$-plane and the 1/2NS-branes are extended
along the $X^{012345}$ directions. Depending on the type of
$Op$-planes, we have different $p$-dimensional gauge theories:
$Sp(k)$ for $Op^+$, $Sp'(k)$ for $\widetilde{Op^+}$, $SO(2k)$ for
$Op^-$ and $SO(2k+1)$ for $\widetilde{Op^-}$. All of them have $n$
hypermultiplets, except for $Sp'(k)$ gauge theory which has two
extra half-hypermultiplets. We can Higgs the above theories down
to, for example, $Sp(k-1), Sp(k-1), SO(2k-1)$ and $SO(2k)$ by
breaking one physical $Dp$-brane between the NS-branes and the
$D(p+2)$-branes. Now we need to match the degrees of freedom
before and after Higgsing. From the field theory point of view,
when we Higgs the gauge theory, one vector multiplet always eats
one hypermultiplet. This is true in any dimension, because there
are 8 supercharges
    \footnote{This fact can be easily seen by noticing that one massive
    vector multiplet has the same number of degrees of freedom
as one massless vector multiplet and one massless hypermultiplet}.
So, after Higgsing,
we get the lower rank gauge theories, some hypermultiplets
in the fundamental representation and some hypermultiplets in the singlet
representation.
From the brane setup point of view, the Higgsing simply corresponds to
the breaking of
$Dp$-branes between 1/2NS-branes and 1/2$D(p+2)$-branes. Matching
    \footnote{When we do the matching, we need to know the generalized
    supersymmetric configuration. This is given in the Appendix. For
    more details, the reader may refer to \cite{Mir}
where one explicit example is given.} the degrees of freedom that
can be read from the brane setup with field theory, we find the
results given at the end of the last paragraph.

\par
To make the above argument more transparent, let us consider one
example, i.e., the splitting of a $D(p+2)$-brane on the
$\widetilde{Op^-}$-plane. The brane setup is given in Figure
\ref{fig:Opminustilde}. The gauge theory is $SO(2k+1)$ with $N$
hypermultiplets (see part (a)). After splitting the
$D(p+2)$-brane, a $Dp$-brane is generated as in part (b). Then we
move the 1/2$D(p+2)$-branes across the two 1/2NS-branes. From part
(c), we see that the remaining theory is $SO(2k)$ with $N-1$
hypermultiplets and $N$ singlets
    \footnote{One easy way to see that there are $N$ singlets is that
    $N-1$ $D(p+2)$-branes will cut the generated $Dp$-brane into
    $N$ pieces.}.
Notice that the $Dp$-brane generated by the splitting of the
$D(p+2)$-brane does not contribute any vector multiplet. Now, let
us count the various degrees of freedom. Before Higgsing, we have
$N(2k+1)=2kN+N$ degrees of freedom. After Higgsing, we have
$(N-1)2k+N=2kN+N-2k$ only. However, taking into account the
$k(2k+1)-k(2k-1)=2k$ degrees of freedom which have been eaten by
the vector multiplets, the two numbers match exactly,
$2kN+N-2k+2k=2kN+N$. The same argument can be applied to  other
cases.
\EPSFIGURE[ht]{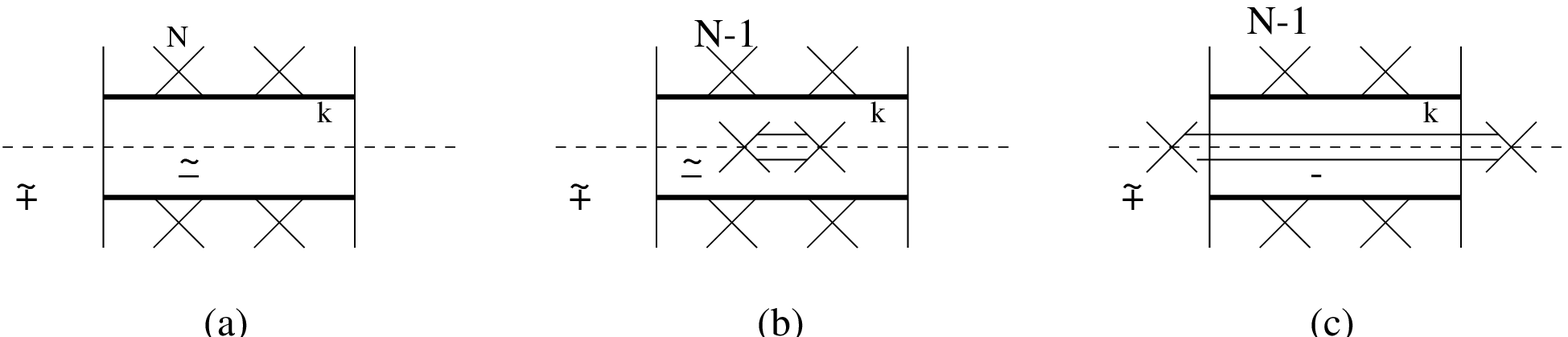,width=14cm}{ (a) The brane setup of
the gauge theory. (b) The splitting of a $D(p+2)$-brane on the
$\widetilde{Op^-}$-plane with the generation of one $Dp$-brane.
(c) The final theory after Higgsing. \label{fig:Opminustilde}}

\par
The second argument is based on T-duality. Let us give an example,
namely the derivation of the splitting of a $D6$-brane on
$\widetilde{O4^-}$, to illustrate the idea. We start from the
known result of the splitting of a $D5$-brane on
$\widetilde{O3^-}$. First, we compactify the $X^3$ direction and
get two fixed $O3$-planes which are shown in part (a) of Fig.
\ref{fig:split}. It is important to notice that the two
$1/2D5$-branes intersect only the upper $O3$-plane. Notice also
that, in order to get $\widetilde{O4^-}$ after T-duality, we set
the lower $O3$-plane to be $O3^-$. After T-duality,
$\widetilde{O3^-}$ and $O3^-$ combine to give $\widetilde{O4^-}$,
while $O3^-$ and $O3^-$ combine to give an $O4^-$, as in part (b).
>From part (b), we can read off that a physical $D4$-brane is
generated when a physical $D6$-brane splits on $\widetilde{O4^-}$.
We can also reach part (b) by splitting one $D5$-brane on the
lower $O3^-$-plane, as shown in part (c). Since two
$\widetilde{O3^-}$-planes yield one $O4^-$ with one physical
$D4$-brane, we obtain the same result.

\par
The T-duality argument can be easily generalized to other cases.
One will find the universal behavior we claimed before. For
instance, we can compactify the $X^2$ direction instead of the
$X^3$ direction to find the splitting of a $D4$-brane on
$O2$-planes. In this case, it is important to recall the result of
\cite{Han1} that, after T-duality, the $O3$-plane will be divided
into two $O2$-planes, but the $1/2D4$-branes will intersect only
one $O2$-plane.
\EPSFIGURE[ht]{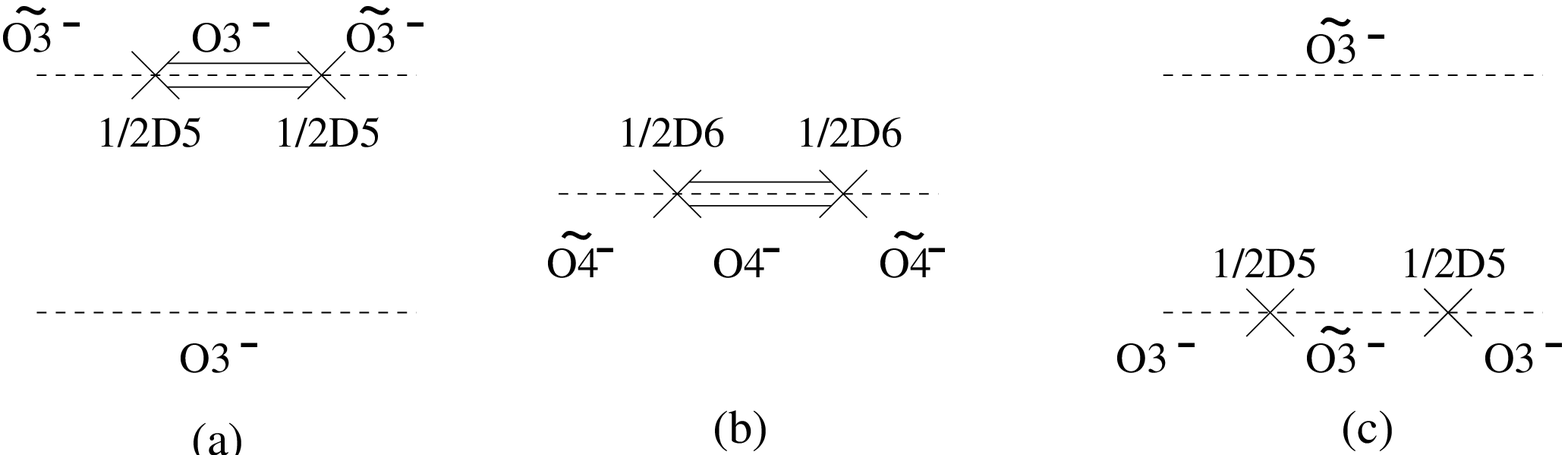,width=14cm}{Part (a) shows that a $D3$-brane
is generated when a $D5$-brane splits on the upper
$\widetilde{O3^-}$-plane. Part (b) is T-dual to Part (a). (c)
Instead of splitting on the upper plane, we split the $D5$-brane
on the lower $O3^-$ plane. No $D3$-branes appear and after
T-duality we recover Part (b) again. \label{fig:split}}

There are other universal behaviours, like the bending of the
branes and the charge difference of the orientifold planes on the
two sides of a 1/2$D(p+2)$-brane. They will be addressed in next
subsection when we discuss the splitting of NS-branes.

All of this evidence supports the claim that the splitting of
$D(p+2)$-branes on $Op$-planes is universal.

\subsection{The non-universal behaviour of the NS-brane splitting}
\label{2.2}

Now let us discuss the splitting of physical NS-branes on
$Op$-planes. One may think that we can solve this problem by
T-duality, but there are actually several observations that
indicate the failure of naive T-duality. For example, it is well
known that a $D4$-brane is created after an NS-brane crosses a
$D6$-brane. But after T-duality, there is no brane creation after
a Kaluza-Klein monopole crosses a $D5$-brane, as shown in
\cite{0103098}.

Let us recall that the NS-branes are extended along the
$X^{012345}$ directions and that the $Op$-plane is
extended along the $X^{012 \ldots (p-1)}$ and $X^6$ directions.

First of all, let us consider the bending of an NS-brane due to
the intersection with $Dp$-branes. From Witten's work \cite{Wit},
such a bending can be found approximately by solving
\begin{equation}
\label{bending}
\bigtriangledown^2 x^6= q \delta^{(6-p)} (r-r_0)
\end{equation}
where function $x^6(r)$ represents the bending of the  NS-brane
along the $X^6$ direction, $q$ is the $D$-brane charge deposited
at the intersection points $r_0$ and $\bigtriangledown^2$ is the
Laplacian operator in the NS-brane worldvolume coordinates. Far
away from the $Op$-plane, the bending of $x^6$ goes like
$r^{p-4}$. So, when $p\leq 3$, the bending caused by the
intersection will disappear at large distance, while when $p=4,5$
the bending will propagate to infinity. This transition from $p=3$
to $p=4$ is the first sign that we should be careful while
applying T-duality.

\par
Let us compare the above bending of NS-branes with the one of
$D(p+2)$-branes. Since a $D(p+2)$-brane always has three
non-compact transverse dimensions relative to the Dp-brane, its
bending goes like $r^{-1}$ and disappears at infinity, which does
not depend on $p$. This difference between the non-universal
bending of NS-branes and the universal bending of $D(p+2)$-branes
indicates why T-duality is applicable to the $D$-brane system but
not to the NS-brane system.

\par
The second observation is related to the charge difference between
the two $Op$-planes on the two sides of the intersection. On one
hand, at the intersection with a 1/2NS-brane, the charge
difference is $2\times 2^{p-5}=2^{p-4}$ for the $Op^{\pm}$ pair
and $2^{p-5}-(\frac{1}{2}-2^{p-5})=2^{p-4}-\frac{1}{2}$ for the
$\widetilde{Op^{\pm}}$ pair, both depending on $p$. On the other
hand, at the intersection with a 1/2$D(p+2)$-brane, the charge
difference is always $\frac{1}{2}$ for the $Op^-,\widetilde{Op^-}$
pair and zero for $Op^+,\widetilde{Op^+}$ pair, no matter what $p$
is.

\par
The third observation concerns the different behaviour of
NS-branes and $D$-branes under T-duality. If we T-dualize one
dimension down, we get two fixed $O(p-1)$-planes. A 1/2NS-brane
will intersect both fixed planes, while a $1/2D(p+1)$-brane (the
T-dual of a $1/2D(p+2)$-brane) will intersect only one of these
two fixed planes. These different intersection properties have
very important consequences. For example, if T-duality is correct,
the $D3$-brane generated on the $O3^+$-plane by splitting the
NS-brane should be divided into two equal parts on both
$O2^+$-planes, i.e. there is a $1/2D2$-brane generated when an
NS-brane splits on an $O2^+$-plane, which is not allowed. If we
went further, we would come up with the conclusion that there is a
quarter physical $D1$-brane generated when an NS-brane splits on
an $O1^+$-plane. This is obviously wrong. In the $D$-brane case,
we do not have this problem because the $D$-branes will intersect
with and split on only one of these two fixed planes, hence we do
not need to divide the effect into two parts.

\par
On these grounds, it is clear we should rely on other methods to
discuss the splitting of NS-branes on $Op$-planes. In the case of
$O3$-planes, we can apply S-duality. For $p\leq 4$, we can use a
minimum energy argument. For $p=5$, we need to analyze the
relative $(p,q)$-web. For $p=6$, the configuration has to be
consistent with RR charge conservation in a background with a
possibly non-vanishing cosmological constant. The details of these
methods are given in the following subsections.

\subsection{The minimum energy argument}\label{minimum}

There are two sources of energy costs when we try to split a
physical brane into two half branes (two 1/2$D(p+2)$-branes or
1/2NS-branes) on the $Op$-plane. The first one is the energy it
takes to bend a 1/2$D(p+2)$-brane or 1/2NS-brane. The bending
energy is proportional to the absolute value of the $Dp$-brane
charge difference between the two sides. The second one is due to
the creation (which costs energy) or annihilation (which produces
energy) of $Dp$-branes, parallel to the $Op$-plane, between the
two 1/2-branes. The minimum configuration is the balance between
these two factors. For $p\leq 4$, since the 1/2NS-brane has more
transverse directions than the $Dp$-branes, the energy cost from
the first source dominates. For $p=5,6$ the second source becomes
important so we need to be more careful. However, in this
subsection, we focus on the $p\leq 4$ case only.

\subsubsection{The $D(p+2)$-brane case}
To check that our minimum energy argument is right, we apply it to
the splitting of $D(p+2)$-branes. Since $D(p+2)$-branes have two
transverse directions more than $Dp$-branes, the first source is
more important. The charge difference between the two sides of the
split 1/2$D(p+2)$-brane is given by the following table
$$
\begin{tabular}{|c|c|}  \hline
Type  &  $\Delta  q$ ~~for~~D(p+2)  \\ \hline
$Op^+$  &   $N$  \\  \hline
$\widetilde{Op^+}$ & $ N$  \\  \hline
$Op^-$  &   $\frac{1}{2}+N$  \\  \hline
$\widetilde{Op^-}$ & $-\frac{1}{2}+N$  \\  \hline
\end{tabular}
$$
where $N$ is the number of $Dp$-branes generated in the middle
when we split the $D(p+2)$-brane. Note that there is no
dependence on $p$, signalling a universal behaviour
for $D(p+2)$-branes.
Thus we see that, in order to minimize the energy cost,
we need to choose $N=0$ in the
first two cases, $N=0,-1$ in the third case and $N=0,1$ in the last case.

We need a complementary argument to fix the ambiguity in the last
two cases. This is provided by field theory as we discussed above.
Namely, we need to match the Higgs branch before and after the
Higgsing mechanism.

\par

The above discussion is applied to all $p \leq 6$, and in
particular also to the $\widetilde{O6^-}$-plane, whose existence
was argued by \cite{Ima}. Note that, upon crossing a 1/2NS-brane,
the $\widetilde{O6^-}$-plane becomes a $\widetilde{O6^+}$-plane
\cite{EJS,Wit1}, which indicates the existence of the
$\widetilde{O6^+}$-plane. Furthermore, since $O6^{\pm}$-planes can
exist in a background with vanishing cosmological constant, they
can also exist with an arbitrary {\sl integer} cosmological
constant. This is due to the fact that, by crossing an arbitrary
number of physical $D8$-planes, the $O6^\pm$-planes remain such
while the cosmological constant jumps by an integer. Conversely,
since we have found that a physical $D8$-brane can split into two
1/2$D8$-branes, we see that the $\widetilde{O6^{\pm}}$-planes can
exist in a background with arbitrary {\sl half-integer}
cosmological constant \cite{Ima}. This observation will be
important later when we discuss the splitting of NS-branes.

\subsubsection{The NS-brane case}\label{result}
Now we calculate the bending energy of a 1/2NS-brane, which for
$p\leq 4$ is proportional to the $Dp$-brane charge difference
between the two sides, which is given by
$$
\begin{tabular}{|c|c|}  \hline
Type  &  $\Delta  q$ ~~for~~NS  \\ \hline
$Op^+$  &   $-2^{p-4}+N$  \\  \hline
$\widetilde{Op^+}$ & $-2^{p-4}+\frac{1}{2}+ N$  \\  \hline
$Op^-$  &   $2^{p-4}+N$  \\  \hline
$\widetilde{Op^-}$ & $-\frac{1}{2}+2^{p-4}+N$  \\  \hline
\end{tabular}
$$
where $N$ is the number of $Dp$-branes generated in the middle
when we split the NS-brane. Note that the result depends on $p$,
giving evidence of non-universal behavior. The number $N$
corresponding to the minimum energy cost is listed below \be
\begin{tabular}{|c|c|c|c|c|}  \hline
Type           & $p=1$ & $p=2$ & $p=3$ & $p=4$ \\ \hline
$Op^+$         & $0$   & $0$  & $0,1$ & $ 1$  \\  \hline
$\widetilde{Op^+}$ & $0$ & $0$ & $0$ & $0,1$   \\  \hline
$Op^-$  &       $0$ & $0$ & $0,-1$ & $-1$  \\ \hline
$\widetilde{Op^-}$ & $0$ & $0$ & $0$ & $0,-1$   \\ \hline
\end{tabular}
\label{resambiguous}\ee

Except for $p=1,2$, the minimum energy requirement does not fix
the number uniquely. We need to discuss each case separately.
For $p=3$, S-duality tells us that $N=1$ for $O3^+$ and $N=0$ for $O3^-$.

\par
For $p=4$, we need to take into account that the bending is like $\ln
r$, so that when two 1/2NS-branes bend towards each other, they
will finally meet and combine to give a physical NS-brane which is
not bent. To see it clearly, let us discuss the example of
$\widetilde{O4^+}$. If we choose $N=1$ for $\widetilde{O4^+}$, the
distance between these two 1/2NS-branes will become infinite as we
move away from the intersection point with the $O4$-plane (see
part (a) of Figure \ref{fig:bending}). Such a bending will cost an
infinite amount of energy, and we can rule it out. On the other
hand, if we choose $N=0$ for $\widetilde{O4^+}$, these two
1/2NS-branes will meet each other (see part (b) Figure
\ref{fig:bending}) and combine to give an NS-brane. This
configuration definitely costs less energy, so we should choose
$N=0$. By the same argument, we have that $N=-1$ for the case
$\widetilde{O4^-}$.

\EPSFIGURE[ht]{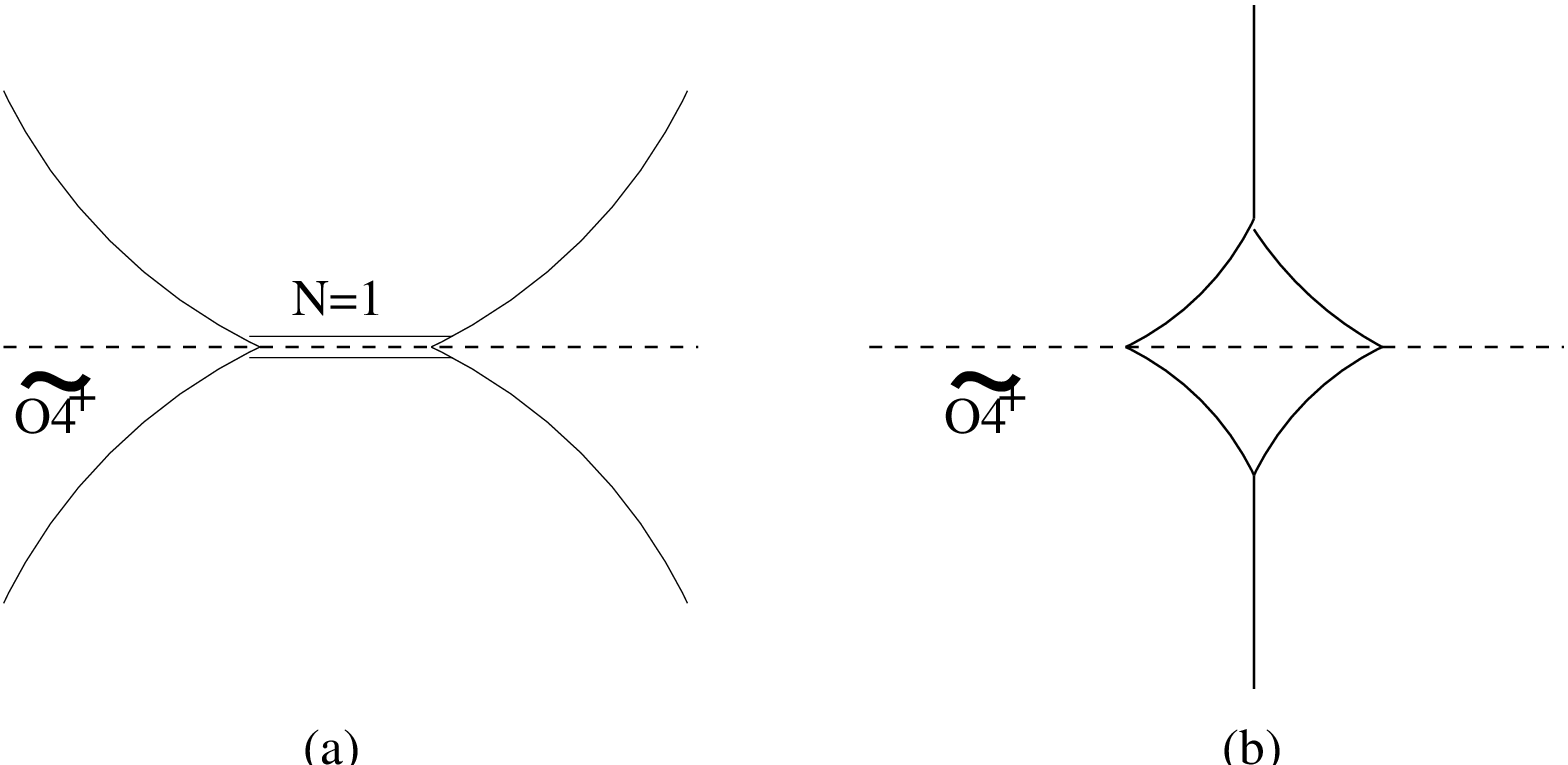,width=10cm} {The bending of 1/2NS-branes
after splitting on the $\widetilde{O4^+}$-plane for case (a) $N=1$
and (b) $N=0$. \label{fig:bending}}

\par
\subsection{The splitting of NS-branes on $O6$-planes}\label{O6}
When we consider the NS/$D6$-brane system, we need to take the
cosmological constant $M$ into account \cite{9706047,9712145}. In
this background, RR charge conservation requires that
$$
N_L-N_R = M\,,
$$
where $N_L$ is the number of $D6$-branes ending on the left hand
side of the NS-brane and $N_R$ is the number of $D6$-branes ending
on the right hand side. In the presence of an orientifold, the
same formula holds for a 1/2NS-brane with the understanding that
$N_L, N_R$ are the D6-brane charges.

The general framework for our discussion is depicted in Figure
\ref{fig:O6split}. In part (a), we have $N_L$ and $N_R$
$1/2D6$-branes ending on the two sides of two 1/2NS-branes and the
consistency condition reads
$$
1/2(N_L-N_R)=M\,.
$$
After the splitting, we have $2N_L$ $1/2D6$-branes ending on the left,
$2N_R$ $1/2D6$-branes ending on the right
and $2N$ $1/2D6$-branes in the middle (see part (b)).
The consistency conditions become
$$
[(N_L+q_1) -(N+q_2)]=M,~~~~~~~~~~[(N+q_2)-(N_R+q_1)]=M
$$
where $q_1, q_2$ are the  charges of the left (or right) and
middle orientifold planes respectively. Notice that these two
conditions imply that $1/2(N_L-N_R)=M$, which is consistent with
part (a). Now we can solve these two equations and get
$$
N_L-N = M + q_2 - q_1
$$
Putting different $q_1,q_2$, we can summarize the results in the
following table
\\
\TABLE[v]
{\begin{tabular}{|c|c|c|c|c|}  \hline
$q_1$ charge  & $O6^+$ & $O6^-$ & $\widetilde{O6^+}$ & $\widetilde{O6^-}$
\\ \hline
$N$ &   $N_L-(M-4)$  & $N_L-(M+4)$  & $N_L-(M-7/2)$  &  $N_L-(M+7/2)$
\\ \hline
\end{tabular}}

Note that since $N_L,N$ and $N_R$ are integers, the cosmological
constant $M$ has to be an integer in the $O6^\pm$ case and
half-integer in the $\widetilde{O6^\pm}$ case for the splitting to
be possible . This condition on $M$ is perfectly consistent with
our observation at the end of section $2.1$.

\EPSFIGURE[ht]{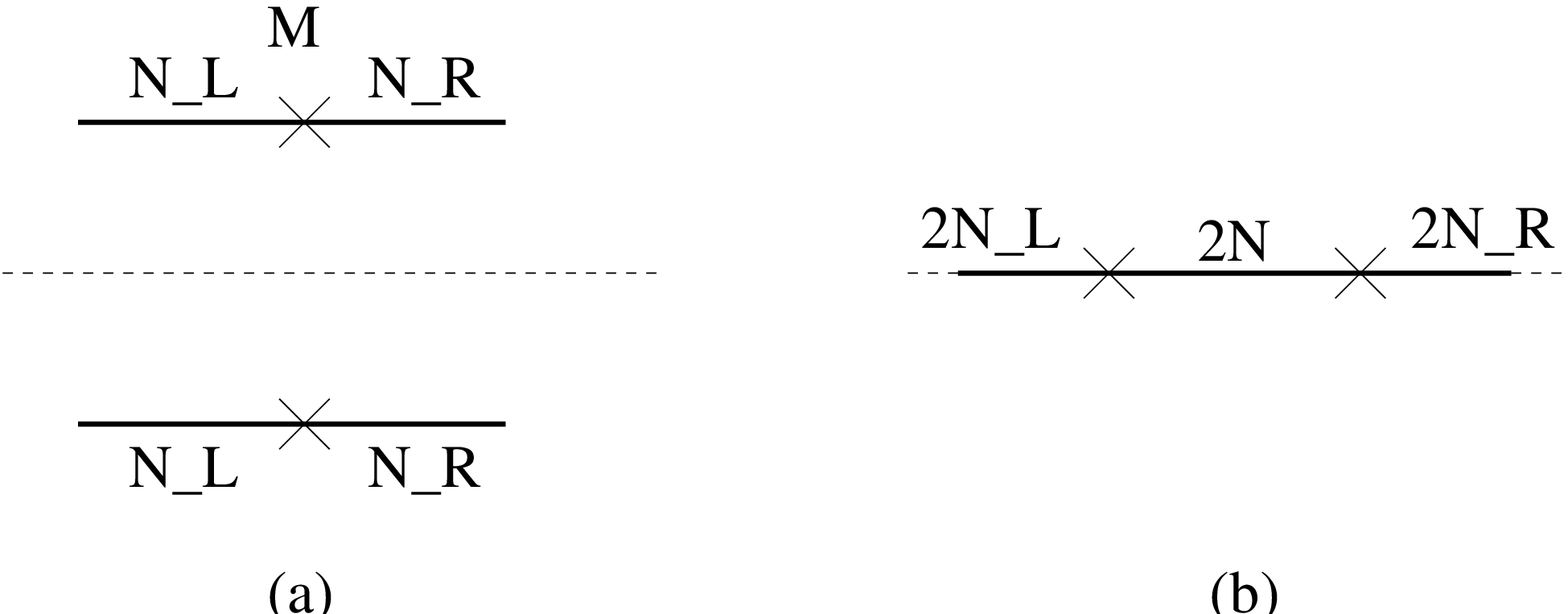, width=14cm}{The splitting of an NS-brane
on $O6$-planes with cosmological constant $M$. (a) Before the
splitting, we have $1/2(N_L-N_R)=M$. (b) After the splitting, $N$
physical (or $2N$ half) $D6$-branes are generated.
\label{fig:O6split}}

\subsection{The splitting of NS-branes on $O5$-planes}\label{O5}
In this case, things become more interesting. We not only have
NS-branes or D5-branes, but also general $(p,q)$-branes with $p,q$
coprime\cite{9705031,9710116}. Charge conservation at the
intersection point becomes the necessary condition for the
configuration to be consistent. Supersymmetry further requires the
orientation of branes to be fixed by their $(p,q)$-charges and the
string coupling constant $g_s$.

Let us recall some facts about the general $(p,q)$-web. Define
\be
\tau=\xi/2\pi +i/\lambda\,.
\ee

The tension of a $(p,q)$-brane is \be T_{p,q}=|p+\tau q|T_{D5}\,,
\ee (from this we see that $(1,0)$ is a $D5$-brane and $(0,1)$ is
an NS-brane. This is the convention used in the paper). To
preserve a quarter of the original supersymmetry, the orientation
of the $(p,q)$-brane is given by \be \Delta x +i \Delta y ||
p+\tau q\,. \ee When we include $O5$-planes along the
$x$-direction, to be consistent with the $Z_2$ orientifold action,
$\tau$ must be a pure imaginary number, namely $\xi=0$. If we use
this condition, we will have that \be T_{p,q}= T_{D5}
\sqrt{p^2+|\tau q|^2} \ee with the orientation \be \Delta x
/\Delta y = p/(-i \tau q) \ee

\EPSFIGURE[ht]{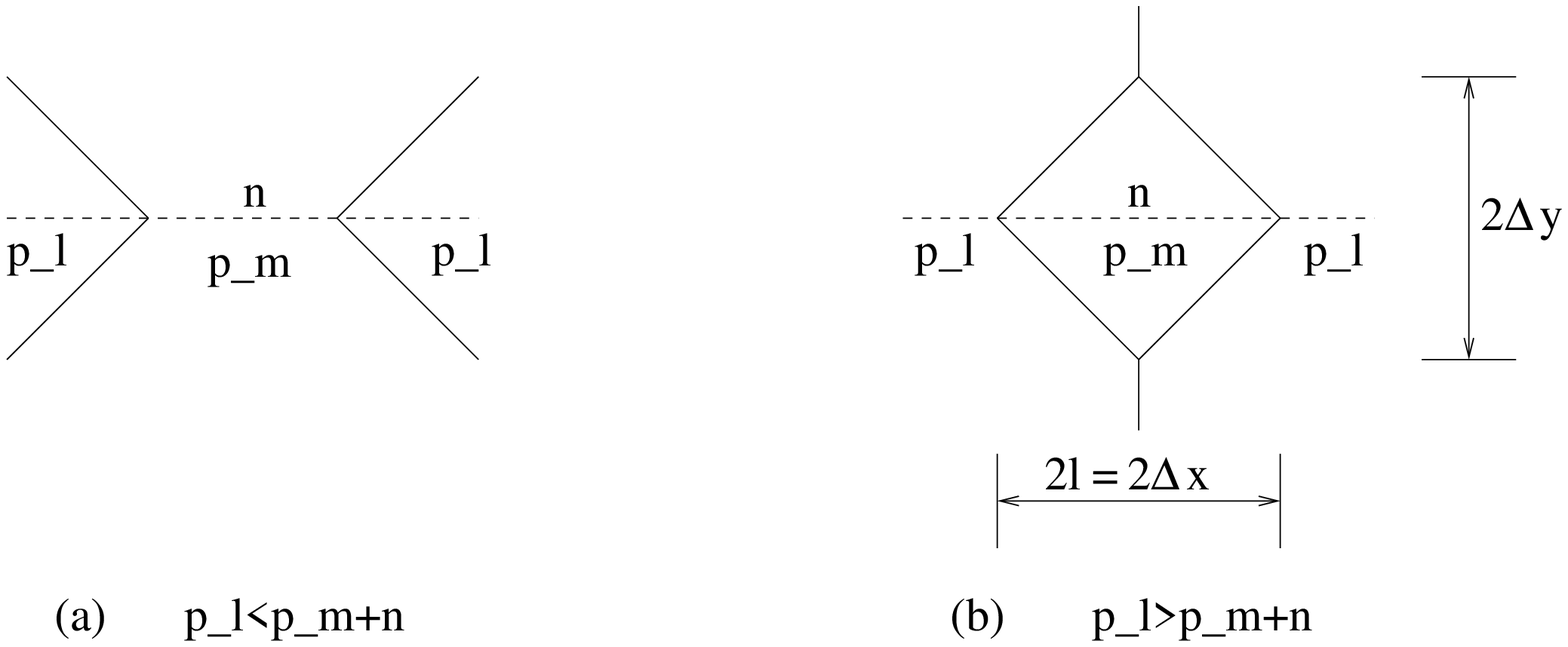, width=14cm}{The setup of NS-brane splitting
on $O5$-planes, where $p_l, p_m$ are the charges of the
orientifolds on the l.h.s. (the same as r.h.s.) and in the middle
and $n$ is the number of $D5$-branes generated in the splitting
process. (a) the bending when $p_l<p_m+n$. (b) the bending when
$p_l>p_m+n$. In this case, the 1/1-branes meet and combine to
become one physical NS-brane again. \label{fig:setup}}

Now we can start the calculation of the energy cost. The general
setup is given in Figure \ref{fig:setup}. When $p_l < p_m+n$, the
brane bending is like in part (a) which will cost an infinite
amount of energy. This leaves us the condition that $p_l \geq
p_m+n$. For $p_l = p_m+n$ with integer $n$, the brane does not
bend and the energy cost is
$$
2l n T_{D5}+ 2l (p_m-p_l) T_{D5}=0
$$
where the first term is the contribution from the generated
D5-branes (remember that negative $n$ means annihilation) and the
second term is the contribution from the  change of orientifold
plane. Next, consider the case $p_l >  p_m+n$ as in part (b) of
Figure \ref{fig:setup}. Charge conservation tells us that the bent
branes are $((p_l - (p_m+n))/2, \pm 1/2)$-branes. From it we get
\bean
|\Delta y| & = & |(\tau l)/(p_l - (p_m+n))|,  \\
length & = & \sqrt{ l^2 + (\Delta y)^2} = l
\sqrt{1+(\frac{|\tau|}{p_l-p_m-n})^2}, \\
E_{bending} & = & 4 T_{p,q} length = lT_{D5} 2(p_l-p_m-n)
    [1+(\frac{|\tau|}{p_l-p_m-n})^2]
\eean
Now the change of energy is
\bean
\Delta E & = & 2l n T_{D5}+2l (p_m-p_l) T_{D5} + E_{bending}- 2  T_{D5}
   |\tau| \frac{ |\tau| l}{ p_l-p_m-n}  \\
& = & 0 \eean where in the first line, the third term (the bending
energy) minus the fourth term (the original non-bending part of
the NS-brane) give the energy cost for the bending.

\par
From this calculation we see that as long as $p_l \geq  p_m+n$,
any $n$ is a good choice since all of them cost the same energy
($zero$). It seems to be a little surprising. However, notice that
we can relate configurations of different $n$ by a smooth motion
with $zero$ energy cost, so that above results make perfect sense.
Now inserting the proper values of $p_l,p_m$, we can find the
results shown in the table in the introduction. For example,
putting $p_l=+1$ ($O5^+$) and $p_l=-1$ ($O5^-$) we get $n\leq 2$.

There are several points we want to discuss. First, we notice that
the change of total energy after splitting on $O5$-planes is $zero$. Does 
this
hold for other $Op$-planes? It does not seem so. 
In general situations, 
the total energy of the system does not change before and 
after the splitting,
but sometimes it does. For example, when an NS-brane splits
on an $O3^-$-plane,
there is no brane created or annihilated, but the 1/2NS-brane
does bend. In other words, in the splitting process we do put energy
into the system.

The change of total energy may cause some confusion because it seems that
during the whole process
the system is supersymmetric and the motion is smooth. In fact, if two
parallel branes are static, they do preserve half the amount of
initial supersymmetry, but if
they move relatively to each other, 
all supersymmetries are broken. The splitting
process, no matter how slow it is, does break supersymmetry, 
so we cannot use
the supersymmetry argument to require that the energy is the same
before and after
splitting. To make this point more clear,
let us recall the following
brane setup given in \cite{Han-Wit}: $N$ D3-branes ending on two
parallel NS-branes. If we keep the positions of the D3-branes on
the NS-branes
invariant and push these two NS-branes away from each other, the
bending of the NS-branes will be same, but the
length of the D3-branes will increase, 
which means that the total energy of the
brane system does increase.

\par
The second point we want to discuss is that for
the $\widetilde{O5^\pm}$ cases, we have D-brane charge
$(p_l-p_m-n)/2=\pm 3/4-n/2$ for the bending brane if we demand the
charge conservation condition at the intersection point. The
appearance of a quarter charge D-brane is a non-trivial prediction by
splitting process. However, the legitimacy of a quarter charge D-brane
is not
very clear to us at this moment. We have several arguments 
to support its existence, but we still cannot draw a   
definite
conclusion from them. First, if we do not allow for 
a quarter charge D-brane, we would violate  
charge conservation at the intersection point and
the energy change before and after splitting would not be zero. 
The second
point is that the D-brane with quarter charge is not complete. It
sits at $X^{789}=0$, like the orientifold plane, and must be
accompanied by its image. If we draw them in real space (not
the covering space in Figure \ref{fig:setup}), they are on top of
each other. So, when we calculate the D5-brane charge, we must sum
them together and we get a half-integer charge which would not violate
the Dirac quantization condition. In comparison, a 1/2D5-brane and
its image can have nonzero $X^{789}$, so when we draw them in 
real space, they are not on top of each other and the D-brane charge
is not the sum.
The third point is that branes with quarter charge
and their images can never be on top of the orientifold because, if they
were, they
would change the type of orientifold and cause an inconsistency.
The last point is that a 1/2NS-brane does exist on
$\widetilde{O5^\pm}$-planes.
This can be seen from the following brane construction given in Figure
\ref{fig:pq1}. In part (a), there are two physical D5-branes ending on
the r.h.s. of a 1/2NS-brane. Then, we move the 1/2D7-brane to the right
and reach
part (b). Since part (a) exists, part (b) must exist too.

\EPSFIGURE[ht]{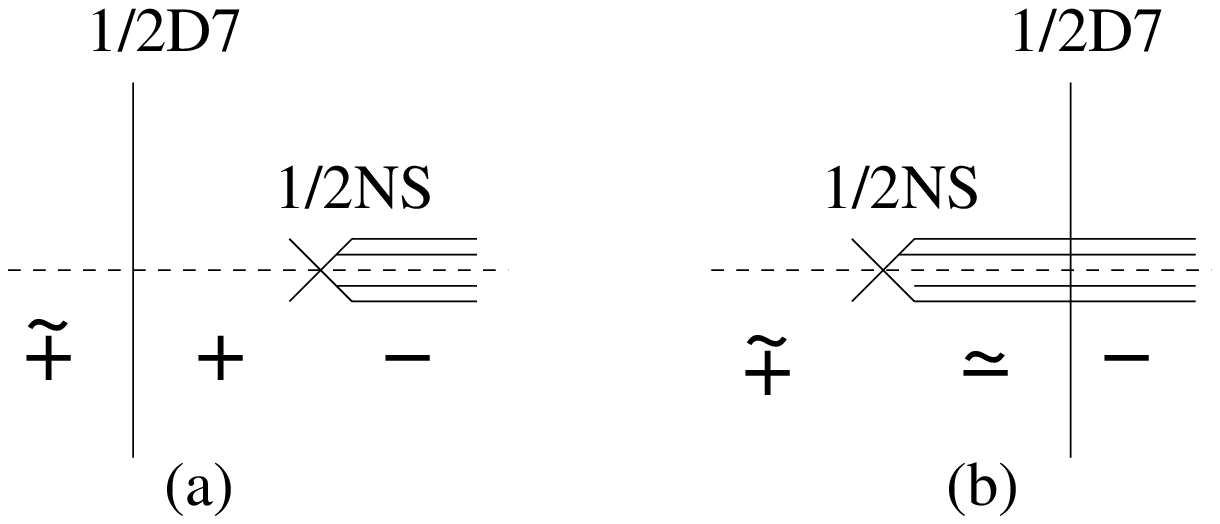, width=14cm}{The proof that a 1/2NS-brane does exist on
$\widetilde{O5^\pm}$-planes. Starting from part (a), which we know is 
an allowed configuration,
by moving the 1/2D7-brane to the right, we reach part (b),
where the 1/2NS-brane
intersects the $\widetilde{O5^\pm}$-plane.
\label{fig:pq1}}

\section{Analysis of $p=4$ by Seiberg-Witten curves}\label{three}
We get the splitting rules in general situations by using various
arguments. However, it would be very satisfactory if we had some
more direct way to derive them. Although we do not know how to do
this for general $p$, in the case of $O4$-planes, we can check the
above results by studying the Coulomb branch of symplectic and
orthogonal gauge theories arising on the worldvolume of
$D4$-branes stretched between the 1/2NS-branes.

The general idea is the following.
Let us start from the brane setup with
$N$ physical $D4$-branes on top of $O4$-planes
and some physical NS-branes above
the $O4$-planes (see, for example, part (a) of Figure \ref{fig:frame}).
Then, we move the NS-branes to touch the $O4$-planes
and split them into 1/2NS-branes, where we assume this is allowed.
Next, we split the $D4$-branes and let
them end on the 1/2NS-branes. After these operations, we get
various gauge theories between the 1/2NS-branes
with rank $N'$ and $N$. As we know $N'$ can differ from $N$.
As a matter of fact, this is exactly the brane configuration
considered in \cite{Land} to derive the Seiberg-Witten curves
for the Coulomb branch of ${\cal N}=2$ symplectic and
orthogonal gauge theories via M-theory, generalizing the work
of Witten for the Coulomb branch of ${\cal N}=2$ $SU(N)$
gauge theories \cite{Wit} (see part (b) of  Figure \ref{fig:frame}).

If the above splitting process is correct,
we can reverse it. Namely,
we can move two 1/2NS-branes towards each other and combine them
into one physical NS-brane which can then leave the $O4$-plane.

The low-energy physics in the Coulomb branch is encoded
in a polynomial with non trivial dependence on the moduli and scales
of the various theories, the Seiberg-Witten curve.
If the reverse process is possible, the Seiberg-Witten curve undergoes
a specific factorization. In the following, we are going to study the
conditions for this transition to happen.

We have to remark that this analysis via the Coulomb branch
is not general, in the sense that the gauge theories
between the 1/2NS-branes have to be asymptotically free
or conformal.
This imposes certain conditions on the number of $D4$-branes
ending on the various 1/2NS-branes.
The condition has a simple geometrical picture.
Since the inverse Yang-Mills coupling, $1/g^2_{YM}$,
is proportional to the distance between the 1/2NS-branes,
in the AF case the distance between them
increases as we move away from the $O4$-plane and
the $D4$-branes, whereas the distance remains invariant
in the conformal case \cite{Wit,Land}.

\subsection{The general setup}
\par
Before delving into a detailed discussion in the next section,
let us set up the
general framework to investigate the splitting of the NS-brane.

We are going to consider only four 1/2NS-branes with $D4$-branes
stretched between them, as in part (c) of Figure \ref{fig:frame}.
In particular, there are no $D4$-branes on the l.h.s of A2
nor on the r.h.s. of C1.
This configuration
corresponds to a product of three gauge theories with a certain number
of flavors for each factor: the type of $O4$-plane
between two 1/2NS-branes dictates the gauge group on the
$D4$-branes worldvolume.
Recall that after crossing a 1/2NS-brane  the minus charged
$O4$-planes are changed into plus charged $O4$-planes and vice versa:
for example, $O4^+$ turns into $O4^-$ and
$\widetilde{O4^-}$ turns into $\widetilde{O4^+}$ \cite{EJS,Wit1}.
In the following table we list the various gauge group
factors from left to right and the number of flavors for each factor
depending on the type of $O4$-plane between A2 and B1,
\be
\label{gensetup}
\begin{tabular}{|c|c|c|}  \hline
Type           &  Gauge groups & Number of Flavors \\ \hline
$O4^+$         & $Sp(N) \times SO(2N') \times Sp(N)$ &
$(N',2N,N')$  \\  \hline
$\widetilde{O4^+}$ & $Sp'(N) \times SO(2N'+1) \times Sp'(N)$ &
$(N'+1,2N,N'+1)$ \\  \hline
$O4^-$  & $SO(2N) \times Sp(N') \times SO(2N)$ & $(N',2N,N')$ \\ \hline
$\widetilde{O4^-}$ & $SO(2N+1) \times Sp'(N') \times SO(2N+1)$ &
$(N',2N+1,N')$ \\ \hline
\end{tabular}
\ee
A careful analysis of the
Seiberg-Witten curve will tell us if one can move the two middle 
1/2NS-branes
on top of each other
by properly adjusting the gauge theory moduli and scales.
Conversely, if this is possible, a physical NS-brane can be split
on the corresponding $O4$-plane.

\EPSFIGURE[ht]{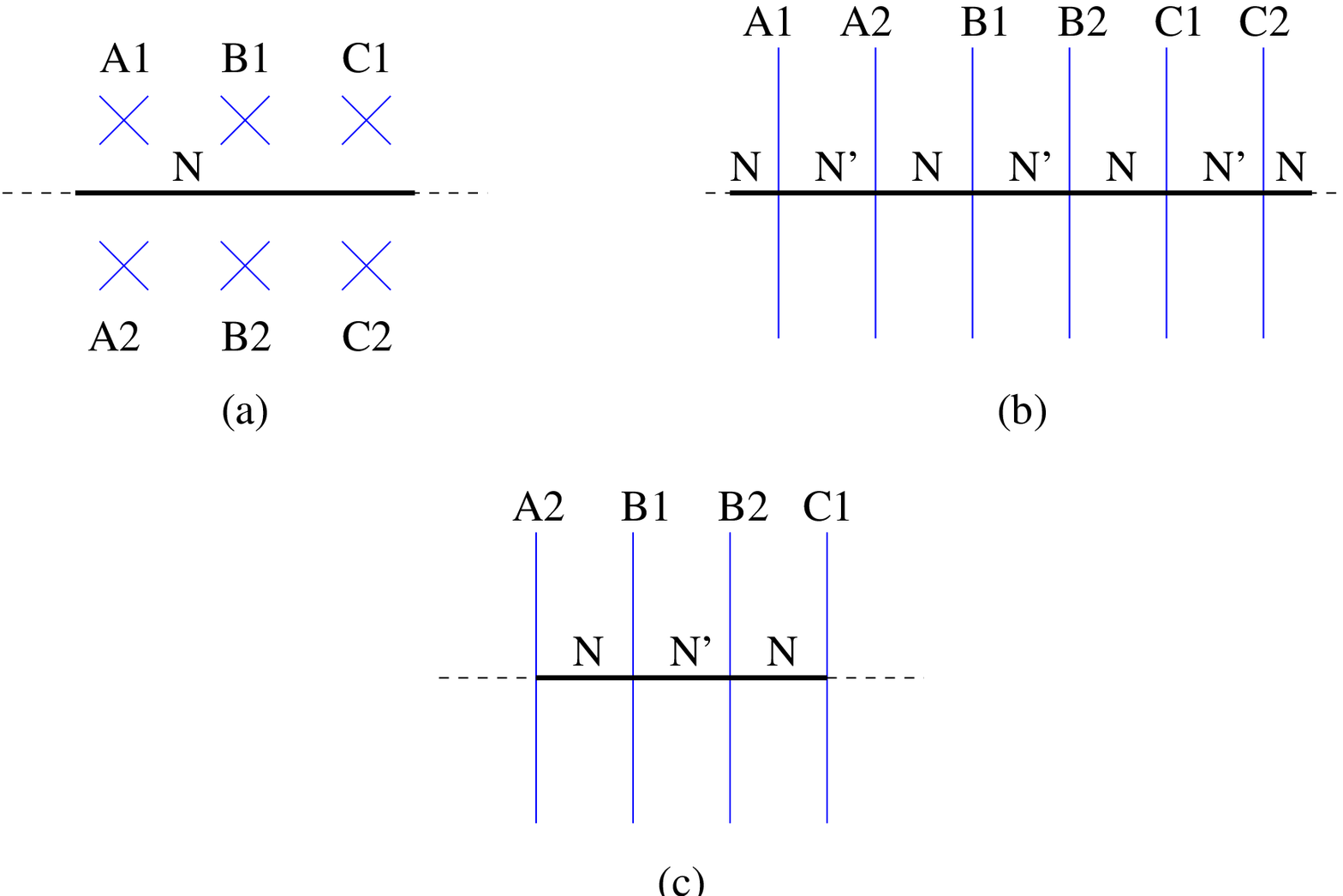,width=14cm}{
(a) Three physical NS-branes before the splitting.
(a) The gauge field theories constructed by the splitting of NS-branes and
$D4$-branes. (c) Our framework of gauge field theories.
\label{fig:frame}}

\subsection{The Seiberg-Witten curve of a single gauge group with flavors}

The results for the ${\cal N}=2$ $SU(2)$ gauge theory obtained
in the celebrated papers \cite{SW1,SW2} have since been
generalized to other gauge groups \footnote{we list only
part of the references and apologize to those who are not cited here}:
to $SU(n)$ groups in \cite{Arg1,Kle,Han22,Arg2}, to
$SO(2k)$ in \cite{Bra,Han3}, to $SO(2k+1)$ in \cite{Han3,Dan}
and to $Sp(k)$ in \cite{Arg,Hoker}.
In this paper, we will follow the conventions of \cite{Arg}.

\subsection{The Seiberg-Witten curves for $SO$ and $Sp$ groups}
The Seiberg-Witten curves for $Sp(r)$, $SO(2r)$ and $SO(2r+1)$ are given
in \cite{Arg}. The coefficients of the $\beta$-functions are
$$
b_{Sp}=2r+2-N_f,~~~~~b_{SO}=N_c-2-N_f~~.
$$
So when $N_f=2r+2$ for $Sp(r)$, $N_f=2r-2$ for $SO(2r)$ and
$N_f=2r-1$ for $SO(2r+1)$, the theories are conformal.
It is understood that the following results apply to asymptotically
free theories only.

The corresponding
curves are summarized in the table

\be
\label{tab_conf}
\begin{tabular}{|l|l|}
\hline
$Sp(r)$ &  $xy^2=(x\prod_{a=1}^r (x-\phi_a^2)+g(\tau)\prod_{j=1}^{2r+2} 
m_j)^2
-g^2(\tau) \prod_{j=1}^{2r+2} (x-m_j^2)$  \\  \hline
$SO(2r)$  & $y^2=x\prod_{a=1}^r (x-\phi_a^2)^2+ 4 f(\tau) x^3 
\prod_{j=1}^{2r-2}(x-m_j^2)$  \\ \hline
$SO(2r+1)$ & $y^2=x\prod_{a=1}^r (x-\phi_a^2)^2+4f(\tau) x^2 
\prod_{j=1}^{2r-1}(x-m_j^2)$  \\  \hline
\end{tabular}
\ee

where
\begin{equation}
\label{g_f}
g(\tau)=\frac{\vartheta_2^4(\tau)}{\vartheta_3^4(\tau)+\vartheta_4^4(\tau)},
~~~~~
f(\tau)=\frac{\vartheta_2^4(\tau) 
\vartheta_4^4(\tau)}{(\vartheta_2^4(\tau)-\vartheta_4^4(\tau))^2}~\,,
\end{equation}
and
$$
\tau = \frac{\theta}{\pi}+ i\frac{8\pi}{g^2}
$$
As we explain in the appendix, following \cite{Arg2} and
\cite{Arg}, $f(\tau+1)$ and $g(\tau+1)$ actually
yield physically equivalent curves.
Matching the brane picture will fix the modular function:
we will encounter this feature when we discuss the splitting of
NS-branes.
Note that sending $\tau$ to $\tau + 1$ amounts
to shifting $\theta$ by $\pi$.

The Jacobi theta functions are defined by
\be
\label{theta}
\begin{array}{rclrcl}
\vartheta_2(\tau) & = & \sum_{n\in Z} q^{(n+\frac{1}{2})^2},~~~~~~~ &
\vartheta_2^4 & = & 16q+O(q^3),
\\ & & & & & \\
\vartheta_3(\tau) & = & \sum_{n\in Z} q^{n^2}, &
\vartheta_3^4 & = & 1+8q+O(q^2),
\\ & & & & & \\
\vartheta_4(\tau) & = & \sum_{n\in Z}(-1)^n q^{n^2}, &
\vartheta_4^4 & = & 1-8q+O(q^2),
\end{array}
\ee
$$
q\equiv e^{i\pi \tau}=e^{i\pi (\frac{\theta}{\pi}+i\frac{8\pi}{g^2})}
=e^{i\theta} e^{-8\pi^2/g^2} \,,
$$
and they satisfy the Jacobi identity
$$
\vartheta_2^4(\tau)-\vartheta_3^4(\tau)+\vartheta_4^4(\tau)=0\,.
$$
Furthermore, in the limit $q\rightarrow 0$,
$$
g(\tau)\rightarrow 8q,~~~~f(\tau)\rightarrow 16 q
$$

\par
To get the curve for the asymptotically free theories with $N_f$ flavors,
for example $N_f<2r+2$ for the $Sp(r)$ group,
we take $2r+2-N_f$ masses to be equal to $M$ and then send
$M$ to $\infty$ while keeping
$\Lambda^{2r+2-N_f}=\Lambda^{b_{Sp}}=qM^{2r+2-N_f}$ finite
\footnote{For the $SO$ group, we should keep
$\Lambda^{2b_{SO}}=qM^{2b_{SO}}$ finite in the large $M$ limit.}
(so $q\rightarrow 0$ in such limit). By such an operation we get
the curve

\be
\label{tab_no_conf}
\begin{tabular}{|l|l|}
\hline
$Sp(r)$ &$ xy^2=(x\prod_{a=1}^r (x-\phi_a^2)+8\Lambda^{b_{Sp}}
\prod_{j=1}^{N_f} m_j)^2
-(-1)^{b_{Sp}}(8\Lambda^{b_{Sp}})^2 \prod_{j=1}^{N_f} (x-m_j^2)$  \\  \hline
$SO(2r)$  & $ y^2=x\prod_{a=1}^r (x-\phi_a^2)^2+(-1)^{b_{SO}}
64\Lambda^{2b_{SO}} x^3 \prod_{j=1}^{N_f}(x-m_j^2)$\\ \hline
$SO(2r+1)$ & $y^2=x\prod_{a=1}^r (x-\phi_a^2)^2+(-1)^{b_{SO}}
64\Lambda^{2b_{SO}} x^2 \prod_{j=1}^{N_f}(x-m_j^2)$  \\  \hline
\end{tabular}
\ee

\subsection{The corresponding curves from M-theory}
The corresponding curves of the above gauge groups, i.e. $Sp(r)$,
$SO(2r)$ and  $SO(2r+1)$ are derived via M-theory in \cite{Land}.
For later purposes, we are interested in finding the explicit
dependence of the various curve parameters on the dynamical scales.

\par
The curves of single gauge groups in \cite{Land} can be summarized as

\be
\label{tab_M_no_conf}
\begin{tabular}{|l|l|}
\hline
$Sp(r)$ & $\prod_{i=1}^{N_f^L}(v^2-m_i^2) t^2+e(v^2B(v^2)-c)t+f
\prod_{j=1}^{N_f^R}(v^2-m_j^2)=0$ \\  \hline
$SO(2r)$  & $v^2 t^2 \prod_{i=1}^{N_f^L}(v^2-m_i^2) + e B(v^2)t+ h v^2
\prod_{j=1}^{N_f^R}(v^2-m_j^2)=0$ \\ \hline
$SO(2r+1)$ &  $v t^2 \prod_{i=1}^{N_f^L}(v^2-m_i^2) +eB(v^2)t+ h v
\prod_{j=1}^{N_f^R}(v^2-m_j^2)=0$  \\  \hline
\end{tabular}
\ee
where $B(v^2)$ is a  polynomial in $v^2$ with degree $r$ and
$e,c,f,h$ are constants which may be functions of the flavour
masses $m_i$ and the dynamical scale $\Lambda$.
Furthermore, the constant $c$ is fixed by requiring the curve
to have a double root at $v^2=0$ \cite{Land}.

\par
Comparing table (\ref{tab_M_no_conf}) with (\ref{tab_no_conf}),
it is easy to find the correspondence

\be
\label{corr}
\begin{tabular}{|l|l|l|}
\hline
group  &  non-conformal   &  conformal  \\  \hline
$Sp(r)$ & $B(v^2) = \prod_{a=1}^r (v^2-\phi_a^2)$  &
  $B(v^2) = \prod_{a=1}^r (v^2-\phi_a^2)$
\\  \hline
& $c  =  -8\Lambda^{b_{Sp}}\prod_{j=1}^{N_f} m_j$ &
$c=-g(\tau) \prod_{i=1}^{2r+2} m_i$
\\  \hline
& $4f/e^2  =  (-1)^{b_{Sp}}(8\Lambda)^{b_{Sp}}$  &
$4f/e^2  = g^2(\tau)$
\\  \hline
$SO(2r)$ & $B^2(v^2)  =  \prod_{a=1}^r (v^2-\phi_a^2)^2$ &
$B^2(v^2)  =  \prod_{a=1}^r (v^2-\phi_a^2)^2$
\\ \hline
& $4h/e^2  =  -(-1)^{b_{SO}}64\Lambda^{2b_{SO}} $ &
$ 4h/e^2  =  -4f(\tau) $
\\ \hline
$SO(2r+1)$ & $B^2(v^2) =  \prod_{a=1}^r (v^2-\phi_a^2)^2$ &
$B^2(v^2) =  \prod_{a=1}^r (v^2-\phi_a^2)^2$
\\ \hline
& $4h/e^2=-(-1)^{b_{SO}}64\Lambda^{2b_{SO}}$  &
$ 4h/e^2  =  -4f(\tau) $
\\ \hline
\end{tabular}
\ee

where we list the results both for the conformal and the
non-conformal case.

\subsection{The $Sp'(r)$ gauge group}
$Sp'(r)$ gauge theory is a little mysterious and there
is no direct result from the field theory point of view.
However, from the  brane setup, it is very natural to write down
the Seiberg-Witten curve. Loosely speaking, $Sp'(r)$ with $N_f$ flavors is
equivalent to $Sp(r)$ with $N_f+1$ flavors, one of which is
massless. Using this correspondence, we can heuristically derive
the curve for $Sp'(r)$.
First, let us recall the $Sp(r)$ case, whose curve is given by
$$
\prod_{i=1}^{N_f^L}(v^2-m_i^2) t^2+e(v^2B(v^2)+8\Lambda^{b_{Sp}}
\prod_{i=1}^{N_f} m_i) t
+ \frac{e^2}{4}(-1)^{b_{Sp}}
64 \Lambda^{2b_{Sp}}
\prod_{j=1}^{N_f^R}(v^2-m_j^2)=0.
$$
To connect with the $Sp'(r)$ case, we multiply by $v$ both the
$t^2$ term and the $t^0$ term. This is because the massless
half-hypermutiplet corresponds to $\sqrt{v^2-m^2}=v$.
Since we have added two half-hypermultiplets,
the number of flavors changes from $N_f$ to $N_f+1$, which implies
$b_{Sp}\rightarrow b_{Sp'}=2r+2-(N_f+1)$. Furthermore,
$$
8\Lambda^{2r+2-N_f}\prod_{i=1}^{N_f} m_i\longrightarrow
8\Lambda^{2r+2-N_f-1}\prod_{i=1}^{N_f+1} m_i=0
$$
because $m_{N_f+1}=0$. Combining all these facts together
we find the curve
\be
\label{Sp_prime}
v \prod_{i=1}^{N_f^L}(v^2-m_i^2) t^2
+e v^2 B(v^2) t+\frac{e^2}{4}(-1)^{b_{Sp'}}
64 \Lambda^{2b_{Sp'}}
v\prod_{j=1}^{N_f^R}(v^2-m_j^2)=0
\ee
with $N_f^L+N_f^R=N_f$. We can factor out one $v$ and get
\be
\label{Sp_prime_1}
\prod_{i=1}^{N_f^L}(v^2-m_i^2) t^2
+e v B(v^2) t+\frac{e^2}{4}(-1)^{b_{Sp'}}
64 \Lambda^{2b_{Sp'}}
\prod_{j=1}^{N_f^R}(v^2-m_j^2)=0
\ee
In the conformal case, the curve is
\be
\label{Sp_prime_2}
\prod_{i=1}^{N_f^L}(v^2-m_i^2) t^2
+e v B(v^2) t+\frac{e^2}{4} g^2(\tau)
\prod_{j=1}^{N_f^R}(v^2-m_j^2)=0
\ee


\section{The Splitting of the NS-brane}\label{four}
After this introduction, it is time to test the results
on the splitting of NS-branes
on $O4$-planes using the Seiberg-Witten curves.
Our general framework is depicted in figure \ref{fig:frame}.
We will first write down the Seiberg-Witten curves for
three gauge groups, including their dynamical scales,
and then check if the curves allow the splitting.
We define $v=X^4+i X^5$, $s=(X^6+iX^{10})/R$
(where $R$ is the radius of compact $X^{10}$ direction), and $t=e^{-s}$
as in \cite{Wit}\cite{Land}.

\subsection
{The curve of $Sp(k_1)\times SO(2k_2) \times Sp(k_3)$}
The beta-function coefficients are
$$
b_{Sp_1}=2k_1+2-k_2,~~~~~b_{SO_2}=2k_2-2-(k_1+k_3).~~~~~
b_{Sp_3}=2k_3+2-k_2,
$$
The curve without dynamical scales
is given in \cite{Land} by
$$
t^4+(v^2B_1(v^2)+C)t^3+B_2(v^2)t^2+(v^2B_3(v^2)+D)t+1=0\,.
$$
In order to find the explicit dependence on them,
we follow the method used in \cite{Asad}.
If we assign the value $2k_1+2$ to the dimension of $t$,
then every term in the curve has dimension $8k_1+8$.
So, we propose the following curve
$$
t^4+(v^2B_1(v^2)+C)t^3+c_1 \Lambda_1^{2b_{Sp_1}} B_2(v^2)t^2+
c_2 \Lambda_1^{4b_{Sp_1}} \Lambda_2^{2b_{SO_2}}  (v^2B_3(v^2)+D)t
+
$$
\be \label{sp_1}
c_3 \Lambda_1^{6b_{Sp_1}} \Lambda_2^{4b_{SO_2}}
\Lambda_3^{2b_{Sp_3}}=0
\ee
The constants $C,D$  are determined by
requiring that the curve (\ref{sp_1}) has double roots
at $v=0$ \cite{Land}. After some algebra, we find
\be
\label{CD_1}
\begin{array}{rcl}
C & = & \sqrt{4  c_1 \Lambda_1^{2b_{Sp_1}} B_2(0)
-8 (x \sqrt{c_3}) \Lambda_1^{3b_{Sp_1}} \Lambda_2^{2b_{SO_2}}
\Lambda_3^{b_{Sp_3}} }, \\
& & \\
D & = &
C \frac{x \sqrt{c_3}}{c_2} \Lambda_1^{-b_{Sp_1}} \Lambda_3^{b_{Sp_3}}\,,
\end{array}
\ee
where $x$ can be $\pm 1$.

\par
After fixing $C$ and $D$, we solve for the constants $c_1,c_2,c_3$
by taking various decoupling limits of (\ref{sp_1}) and
matching the result with the curves for single gauge groups.
For example, sending first $\Lambda_3$ to zero and then
$\Lambda_2$, we decouple the gauge groups $SO(2k_2)$ and $Sp(k_3)$,
and we are left with the curve
$$
t^2+(v^2B_1(v^2)+C')t+c_1 (\Lambda_1^{2b_{Sp_1}})B_2(v^2) = 0\,.
$$
Matching the above with the curve of $Sp(k_1)$
in Table \ref{tab_M_no_conf}, we find
$$
c_1=16(-1)^{b_{Sp_1}}.
$$
It is a little tricky to determine the constant $c_2$. Again, we decouple
$Sp(k_3)$ by letting
$\Lambda_3\rightarrow 0$. This yields
$$
t^3+(v^2B_1(v^2)+C)t^2+c_1 \Lambda_1^{2b_{Sp_1}} B_2(v^2)t+
c_2 \Lambda_1^{4b_{Sp_1}} \Lambda_2^{2b_{SO_2}}  (v^2B_3(v^2)+D)=0.
$$
To decouple $Sp(k_1)$, we need to rescale $t$ first.
Substituting $t$ with $t\Lambda_1^{2b_{Sp_1}}$ in the above equation,
cancelling the common factor $\Lambda_1^{4b_{Sp_1}}$ and
then letting $\Lambda_1\rightarrow 0$, we finally come to
$$
(v^2B_1(v^2)+C)t^2+c_1 B_2(v^2)t+c_2(\Lambda_2^{2b_{SO_2}})  (v^2B_3(v^2)+D)
=0
$$
In this limit, we also have $C=0,D=0$.
Matching with the $SO(2k_2)$ curve in Table \ref{tab_M_no_conf}, we find
$$
c_2=16c_1^2 (-)^{b_{SO_2}+1}=16^3 (-)^{b_{SO_2}+1}
$$
Repeating the procedure once again, we find
$$
\frac{c_3}{c_1}=16 \left(\frac{c_2}{c_1}\right)^2
(-1)^{b_{Sp_3}},~~~~~
D'=8 \Lambda_3^{b_{Sp_3}} \sqrt{(-1)^{k_2} B_2(0)}\,.
$$
Therefore
\be
\label{3.1.c}
c_1=16(-1)^{b_{Sp_1}},~~~~~c_2=16^3 (-1)^{b_{SO_2}+1},~~~~~~
c_3=16^6 (-1)^{b_{Sp_3}} =16^6 (-1)^{b_{Sp_1}+b_{Sp_3}}\,,
\ee
and $x=(-1)^{b_{SO_2}+1}$.
\par
Note that the curve (\ref{sp_1})
is symmetric under the transformation $t\rightarrow t$ and
$v\rightarrow -v$. This is a consequence of the orientifold action,
which reads
$X^{6,10} \rightarrow X^{6,10}$ and $X^{4,5}\rightarrow -X^{4,5}$
and is equivalent to the above transformations of $t$ and $v$ \cite{Land}.

If $b_{SO_2}=0$, the curve is obtained from the above
by replacing $16 (-1)^{b_{SO_2}} \Lambda_2^{2b_{SO_2}}$
with $f(\tau_2)$, where $f(\tau_2)$ is given in (\ref{g_f}),
$$
t^4+(v^2B_1(v^2)+C )t^3
+ 16(-1)^{b_{sp1}}
\Lambda_1^{2b_{Sp_1}}B_2(v^2)t^2
$$
\be
-2^8 \Lambda_1^{4b_{Sp_1}} f(\tau_2)
(v^2B_3(v^2)+ C \Lambda_1^{-b_{Sp_1}} \Lambda_3^{b_{Sp_3}} )t +
2^{16} \Lambda_1^{6b_{Sp_1}} f(\tau_2)^2
\Lambda_3^{2b_{Sp_3}} = 0,
\l{sp_1ciz2}\ee
and
\be
C = 8 \Lambda_1^{b_{Sp_1}} \sqrt{(-1)^{b_{Sp_1}}
B_2(0) + 2^5 \Lambda_1^{b_{Sp_1}} f(\tau_2) \Lambda_3^{b_{Sp_3}} }\,.
\l{ciz2}\ee

\subsection{Lifting of the middle 1/2NS-branes}

Our goal is to understand which conditions the curve (\ref{sp_1})
must satisfy in order to be able to lift the middle
1/2NS-branes, B1 and B2, out of the orientifold plane (see part (c) of
figure
\ref{fig:frame}).
The remaining configuration, with A2 and C1 only, would
describe a single symplectic gauge group theory.


First of all, if we want to move the two 1/2NS-branes B1 and B2 out
of the orientifold plane, we have to reconnect the
$D4$-branes to the left of B1 with the $D4$-branes to the right
of B2.
In particular, the two symplectic gauge groups at both ends
of the configuration must have the same rank.
Thus, we will set $k_1=k_3$ in (\ref{sp_1}),
which leads to $b_{Sp_1}=b_{Sp_3}\equiv b_{Sp}$.

Furthermore, since in this case the left and right gauge theory are
the same, we can write the curve in a more symmetric form which is invariant
under the exchange of the left and right gauge groups.
Recalling that $X^6$ and $X^{10}$ are the coordinates
along the orientifold plane, the above requirement
amounts to the curve (\ref{sp_1}) being symmetric under
$X^{6,10}\to -X^{6,10}$, after setting the origin of
these coordinates appropriately. This transformation
is equivalent to $t \to 1/t$.

This gives a relation between $B_1(v^2)$
and $B_3(v^2)$ in (\ref{sp_1}).
Setting $\Lambda_1=\Lambda_3\equiv \Lambda$, and
performing the rescaling
$t\rightarrow i 2^6 \Lambda^{2b_{Sp}} \Lambda_2^{b_{SO_2}}t
\equiv A t$, the curve becomes
\be
\label{sym_sp_1}
{\cal P}(v^2,t) \equiv
A^4 t^4+ (v^2 B_1(v^2)+C) A^3 t^3+ c_1 \Lambda^{2b_{Sp}} B_2(v^2) A^2 t^2
+(v^2 B_3(v^2)+C)  A^3 t+ A^4 = 0
\ee
where we have used the fact that $C=D$ by (\ref{CD_1}).
Finally, we see that, in order to achieve the
$t \to 1/t$ symmetry,
\be
t^4 {\cal P}(v^2,1/t) = {\cal P}(v^2,t)\,,
\label{mammasonofelice1}\ee
$B_1(v^2)$ must be equal to $B_3(v^2)$.

At a certain stage in the lifting procedure,
the middle NS5-branes will coincide. We want to
argue that in this limit
\be
{\cal P}(v^2,t) \rightarrow {\cal R}(t){\cal Q}(v^2,t),
\label{mammasonofelice2}\ee
where ${\cal R}(t)$ is a quadratic polynomial in $t$ with a
double root and ${\cal Q}(v^2,t)$ is another quadratic
polynomial in $t$ with some $v$ dependence describing
an $Sp(r)$ theory with no flavours.
Recall from \cite{Wit} and \cite{Land}
that the roots in $t$ of ${\cal P}(v^2,t)$ at a given $v$
correspond to the positions of the NS5-branes in the $t$-plane.
Requiring no $v$ dependence and a double root for ${\cal R}(t)$
precisely depicts the situation where the middle
NS5-branes are on top of each other.
Furthermore, we can manipulate (\ref{tab_M_no_conf})
so that ${\cal Q}(v^2,t)$ itself will have a $t \to 1/t$
symmetry
$$
t^2 {\cal Q}(v^2,1/t) = {\cal Q}(v^2,t)\,.
$$

We are now going to show that for the case at hand,
namely when the middle orthogonal theory is not conformal,
the splitting procedure is not possible.

First of all, let us find ${\cal R}(t)$.
Then, Eqs. (\ref{mammasonofelice1}) and
(\ref{mammasonofelice2}) imply that
$$
t^4 {\cal R}(1/t){\cal R}^{-1}(t) =
{\cal Q}(v^2,t){\cal Q}(v^2,1/t)^{-1} =
{\cal Q}(0,t){\cal Q}(0,1/t)^{-1} = t^2,
$$
which fixes ${\cal R}(t)$ to
be proportional to $(t + x)^2$,
where $x = \pm 1$.
Setting
$$
{\cal P}(v^2,t)=
(t+x)^2
\left( 2^8 \Lambda_1^{2b_{Sp_1}}\Lambda_2^{2b_{SO_2}} t^2
+ (e^2 v^2 \tilde{B}(v^2)+ \beta) t + 2^8 \Lambda_1^{2b_{Sp_1}}
\Lambda_2^{2b_{SO_2}}  \right) = 0,
$$
we derive the following relations
$$
- 4i \Lambda_2^{b_{SO_2}}(v^2B_1(v^2)+C) =
e^2 v^2 \tilde{B}(v^2)+ \beta\,,
$$
$$
(-1)^{b_{Sp_1}+1} B_2(v^2) =
2 x  (e^2 v^2 \tilde{B}(v^2)+ \beta) +
2^9 \Lambda_1^{2b_{Sp_1}}\Lambda_2^{2b_{SO_2}}\,.
$$
In particular
$$
(-1)^{b_{Sp_1}+1} B_2(v^2) -
2^9 \Lambda_1^{2b_{Sp_1}}\Lambda_2^{2b_{SO_2}} =
- 8 x i \Lambda_2^{b_{SO_2}}(v^2B_1(v^2)+C)\,,
$$
implying that, in the non trivial case, namely
when $\Lambda_2^{b_{SO_2}}$ does not vanish,
$v^2 B_1(v^2)$ and $B_2(v^2)$ must have the same order.
But this is equivalent to $k_1+1=k_2$ or $b_{SO_2}=0$.
Therefore, we conclude that the lifting cannot happen
when the middle orthogonal theory is not conformal.

Let us turn to the curve (\ref{corr}) describing the
conformal case.
If we set $f(\tau_2)=-\alpha^2$,
with $\alpha$ to be determined,
and perform a redefinition of $t$, $t \to 2^4 \Lambda_1^{2b_{Sp_1}}
\alpha \,t$, the curve becomes
$$
2^{12} \Lambda_1^{6b_{Sp_1}} \alpha^2
\left( 2^4 \alpha^2 \Lambda_1^{2b_{Sp_1}} t^4 +
\alpha (v^2B_1(v^2)+C)t^3
+ \right.
$$
$$
\left.
(-1)^{b_{sp1}} B_2(v^2)t^2 +
\alpha (v^2B_1(v^2)+ C) t +
2^4 \alpha^2 \Lambda_1^{2b_{Sp_1}}
\right) = 0.
$$
In this form, the curve has a $t \to 1/t$ symmetry.
In the limit where the middle NS5-branes coincide,
this should be equal to
$$
2^{12} \Lambda_1^{6b_{Sp_1}} \alpha^2 (t + x)^2
\left( 2^4 \Lambda_1^{2b_{Sp_1}} \alpha^2 t^2
+ (e^2 v^2 \tilde{B}(v^2)+ \beta) t + 2^4 \Lambda_1^{2b_{Sp_1}}
\alpha^2 \right) = 0,
$$
where $x=\pm 1$.
This equation yields
\be
\alpha(v^2B_1(v^2)+ C) = e^2 v^2 \tilde{B}(v^2)+ \beta + 2^5 x \alpha^2
\Lambda_1^{2b_{Sp_1}}\,,
\label{mamma3}\ee
and
\be
(-1)^{b_{Sp_1}} B_2(v^2) = 2 x e^2 v^2 \tilde{B}(v^2) +
2 x \beta
+ 2^5 \alpha^2 \Lambda_1^{2b_{Sp_1}}\,.
\label{mamma4}\ee
Therefore
\be
(-1)^{b_{Sp_1}} B_2(v^2) = 2x\alpha(v^2B_1(v^2)+ C) -
2^5 \alpha^2 \Lambda_1^{2b_{Sp_1}}\,.
\l{uella1}\ee
Since the highest order coefficients both in
$B_1(v^2)$ and $B_2(v^2)$ are $1$, the above equation
implies that $\alpha = {(-1)^{b_{Sp_1}} \over 2}x$
and
\be
f(\tau_2) = -{1 \over 4}.
\label{fff}\ee
Furthermore, using (\ref{ciz2}), where we set
$\Lambda_1^{b_{Sp_1}} = \Lambda_3^{b_{Sp_3}}$,
(\ref{mamma3}) and
(\ref{mamma4}), we get that
$$
\alpha C = 2^6 x \alpha^2 \Lambda_1^{b_{Sp_1}}, \quad
\beta = 2^5 x \alpha^2 \Lambda_1^{b_{Sp_1}}\,.
$$
The remaining symplectic theory is then described by the curve
\be
2^4 \alpha^2 (\Lambda_1^{2b_{Sp_1}}) t^2 +
\alpha(v^2B_1(v^2) + 2^5 x \alpha
\Lambda_1^{2b_{Sp_1}}) t +
2^4 \alpha^2 (\Lambda_1^{2b_{Sp_1}}) = 0,
\label{uella2}\ee
to be compared with
$$
t^2 + e(v^2 \tilde{B}(v^2)+8\Lambda^{b_{Sp}})t
+16 e^2 \Lambda^{2b_{sp}}=0.
$$
Note that $b_{Sp}=2k_1+2=2b_{Sp_1}=2(2k_1+2-k_2)=2(k_1+1)$.
By setting $2^2 \alpha \Lambda_1^{b_{Sp_1}} t \to t$,
(\ref{uella2}) becomes
$$
t^2 +
{1\over 4}
\Lambda_1^{-b_{Sp_1}}(v^2B_1(v^2) + 2^5 x \alpha
\Lambda_1^{2b_{Sp_1}}) t +
2^4 \alpha^2 (\Lambda_1^{2b_{Sp_1}}) = 0,
$$
and everything actually works if we make the
following identifications
\be
\tilde{B}(v^2) = B_1(v^2), \quad
e = {1 \over 4}\Lambda_1^{-b_1}, \quad
\Lambda^{b_{Sp}} = 2 (-1)^{b_{Sp_1}} \Lambda_1^{2b_{Sp_1}}\,,
\l{uella3}\ee
namely
$$
16 e^2 \Lambda^{2b_{sp}} = 2^4 \alpha^2 \Lambda_1^{2b_{Sp_1}}
= 4 \Lambda_1^{2b_{Sp_1}}\,.
$$
In summary, we have seen that we can lift the middle
1/2NS5-branes out of the configuration provided
that $b_{SO_2}=(2k_2-2-2k_1)=0$,
which is equivalent to
$$
k_2 = k_1 + 1\,.
$$
If we look up the $O4^+$-plane in table (\ref{gensetup}),
we see that this corresponds to $N'=N+1$.
Therefore, we reproduced the result listed in
table (\ref{result1}) that a $D4$-brane is created when a physical
NS5-brane splits on an $O4^+$-plane.

\subsection{Consistency check}

Note that, as the two middle 1/2NS-branes approach
one another, the gauge coupling $g_{YM}$ of the
orthogonal theory increases and
$\theta$-angle becomes smaller.
This is because $1/g_{YM}^2$ and $\theta$
are proportional to their relative displacements
$\Delta X^6$ and $\Delta X^{10}$ respectively
\cite{Wit}\cite{Land}.

Therefore, in the limit we just considered,
$g_{YM}$ should diverge and $\theta$ should vanish.
This means that Eq.(\ref{fff}) must have $\tau_2=0$
as a solution.
Using the definition of $f(\tau_2)$ (\ref{g_f}), we find
$$
\vartheta_2^4(\tau_2) \vartheta_4^4(\tau_2) =
-{1 \over 4} {(\vartheta_2^4(\tau_2)-\vartheta_4^4(\tau_2))^2}
\Rightarrow
\left( {1 \over 2} \vartheta_2^4(\tau_2)
+ {1 \over 2} \vartheta_4^4(\tau_2) \right)^2 = 0\,,
$$
and by the Jacobi identity this is equivalent to
$$
\vartheta_3(\tau_2) = 0\,.
$$
The zero points of $\vartheta_3(\tau_2)$ are given by
$$
0={\pi \over 2}+{\pi \over 2}\tau_2 + m \pi + n \pi \tau_2,
~~m,n,\in {\bf Z}
$$
which do not admit $\tau_2 = 0$ as a solution.
However, as we observed above and we explain in Appendix B,
we could select $f(\tau_2 + 1)$ as our modular function,
leading to physically equivalent results.
The condition for the splitting would then become
$\vartheta_3(\tau_2+1) = \vartheta_4(\tau_2) = 0$,
equivalently
$$
0= \frac{\pi}{2} \tau_2 + m' \pi + n' \pi \tau_2,
~~m',n',\in {\bf Z}\,,
$$
which admits $\tau_2 = 0$ as a solution.

\subsection{The curve of $SO(2k_1)\times Sp(k_2)\times SO(2k_3)$}

The beta-function coefficients of the three groups are
$$
b_{SO_1}=2k_1-2-k_2,~~~~b_{Sp_2}=2k_2+2-(k_1+k_3),~~~~~
b_{SO_3}=2k_3-2-k_2
$$
The curve without dynamical scales in \cite{Land} is
$$
v^2 t^4+B_1(v^2) t^3+(v^2 B_2(v^2)+C) t^2+B_3(v^2) t+ v^2=0
$$
Again, if we assign the value $2k_1-2$ to the dimension of $t$,
the curve will be given by
$$
v^2 t^4+B_1(v^2) t^3+c_1 (\Lambda_1^{2b_{SO_1}}) (v^2 B_2(v^2)+C) t^2+
$$
\be
\label{even_1}
c_2 (\Lambda_1^{2b_{SO_1}})^2(\Lambda_2^{2b_{Sp_2}})B_3(v^2) t+
c_3 (\Lambda_1^{2b_{SO_1}})^3(\Lambda_2^{2b_{Sp_2}})^2
(\Lambda_3^{2b_{SO_3}})
v^2=0
\ee
The constant $C$ is determined by requiring that the above equation
has a double root at $v=0$. Therefore
$$
C = \pm 2{\Lambda_2^{b_{Sp_2}} \over c_1}
\sqrt{c_2 B_1(0)B_3(0) }\,.
$$
By decoupling the various groups, we find the coefficients to be
\be
\label{even_c}
c_1=16 (-1)^{b_{SO_1}+1},~~~~c_2=16^3 (-1)^{b_{Sp_2}},
~~~~~c_3=16^6 (-1)^{b_{SO_1}+1+b_{SO_3}+1} = 16^6\,.
\ee

\par Finally, let us consider the conformal case, $b_{Sp_2}=0$,
where $2k_2+2 =k_1+k_3$. In this case, we replace
$8 \Lambda_2^{2k_2+2-k_1-k_3}$ with $g(\tau_2)$, as in
(\ref{g_f}), and rewrite (\ref{even_1}) as
$$
v^2 t^4+B_1(v^2) t^3+c_1 (\Lambda_1^{2b_{SO_1}}) (v^2 B_2(v^2)+C) t^2+
$$
\be
\label{even_2}
c_2 (\Lambda_1^{2b_{SO_1}})^2 g^2(\tau_2)B_3(v^2) t+
c_3 (\Lambda_1^{2b_{SO_1}})^3 g^4(\tau_2)
(\Lambda_3^{2b_{SO_3}})
v^2=0
\ee
where
\be
c_1=16 (-1)^{b_{SO_1}+1},~~~c_2=64,~~~~
c_3=16^3\,,
\label{cccc}\ee
and
\be
C= g(\tau_2)\sqrt{(-1)^{k_1+k_3}B_1(v^2=0) B_3(v^2=0)}\,.
\label{CCCC}\ee

We then proceed to discuss the lifting of the middle
1/2NS-branes.
Setting $k_1=k_3$, $B_1(v^2)=B_3(v^2)$, $\Lambda_1^{2b_{SO_1}}
= \Lambda_3^{2b_{SO_3}}$, and performing the rescaling
$t \to 2^6 \Lambda_1^{2b_{SO_1}}\Lambda_2^{b_{Sp_2}} i t
\equiv A t$,
the curve (\ref{even_1}) becomes
\be
{\cal P}(v^2,t) \equiv
A^4 v^2 t^4 + B_1(v^2) A^3  t^3
+ c_1 \Lambda_1^{2b_{SO_1}} (v^2B_2(v^2)+C) A^2 t^2
- B_1(v^2) A^3 t + A^4 v^2 = 0\,.
\label{uellala33}\ee
The curve ${\cal P}(v^2,t)$ has
a $t \to -1/t$ symmetry
$$
t^4 {\cal P}(v^2,-1/t) = {\cal P}(v^2,t)\,.
$$
In the limit where the middle NS5-branes are lifted,
${\cal P}(v^2,t)$ reduces to
$$
{\cal P}(v^2,t) = {\cal R}(t){\cal Q}(v^2,t),
$$
where the curve ${\cal Q}(v^2,t)$ describes an $SO(2r)$
curve with no flavours.
We have that
$$
t^4 {\cal P}(v^2,-1/t) = t^4 {\cal R}(-1/t) {\cal Q}(v^2,-1/t)
= {\cal P}(v^2,t) = {\cal R}(t){\cal Q}(v^2,t)\,
$$
is solved by ${\cal R}(t)=(t+xi)^2$, where $x =\pm 1$.
Therefore, setting
$$
{\cal P}(v^2,t) = (t + x i)^2
\left( 2^8 \Lambda_1^{2b_{SO_1}}\Lambda_2^{2b_{Sp_2}} v^2 t^2 +
\alpha \tilde{B}(v^2) t
-2^8 \Lambda_1^{2b_{SO_1}}\Lambda_2^{2b_{Sp_2}} v^2 \right)\,,
$$
implies the following identities
$$
- 4 i\Lambda_2^{b_{Sp_2}} B_1(v^2) =
\alpha \tilde{B}(v^2) + 2^9 x i
\Lambda_1^{2b_{SO_1}}\Lambda_2^{2b_{Sp_2}} v^2,
$$
$$
(-1)^{b_{SO_1}} (v^2B_2(v^2)+C) =  - 2^9
\Lambda_1^{2b_{SO_1}}\Lambda_2^{2b_{Sp_2}} v^2
+ 2 x i \alpha \tilde{B}(v^2)\,.
$$
In particular we get
$$
(-1)^{b_{SO_1}} (v^2B_2(v^2)+C)+
2^9 \Lambda_1^{2b_{SO_1}}\Lambda_2^{2b_{Sp_2}} v^2 =
8 x \Lambda_2^{b_{Sp_2}} B_1(v^2) +
2^{10} \Lambda_1^{2b_{SO_1}}\Lambda_2^{2b_{Sp_2}} v^2\,.
$$
In the non trivial case, $\Lambda_2^{b_{Sp_2}}
\ne 0$, the order of the two polynomials on either
side of the equation
is the same and therefore $b_{Sp_2}=2k_2+2-2k_1=0$.
Hence, it is not possible to lift the middle NS5-branes
when the middle symplectic theory is not conformal.

In the conformal case, we obtain
$$
{\cal P}(v^2,t) = \left( 2^3 \Lambda_1^{2b_{SO_1}} g^2 v^2 t^4
-i g B_1(v^2) t^3  +2(-1)^{b_{SO_1}}(v^2 B_2(v^2)+C)t^2
+ \right.
$$
$$
\left.
ig B_1(v^2) t + 2^3 \Lambda_1^{2b_{SO_1}} g^2 v^2 \right)\, =0\,.
$$
Setting
$$
{\cal P}(v^2,t) = (t + x i)^2
\left(  2^3 \Lambda_1^{2b_{SO_1}} g^2  +
\alpha \tilde{B}(v^2) t
- 2^3 \Lambda_1^{2b_{SO_1}} g^2 v^2  \right)\,
$$
we find that
$$
-i g B_1(v^2) =  \alpha \tilde{B}(v^2) +
2^4 x i  \Lambda_1^{2b_{SO_1}} g^2 v^2 \,,
$$
and
$$
2(-1)^{b_{SO_1}}(v^2 B_2(v^2)+C) =
- 2^4 \Lambda_1^{2b_{SO_1}} g^2 v^2
+ 2 x i \alpha \tilde{B}(v^2).
$$
Therefore
$$
2(-1)^{b_{SO_1}}(v^2 B_2(v^2)+C) +
2^4 \Lambda_1^{2b_{SO_1}} g^2 v^2
= 2x g B_1(v^2) + 2^5 \Lambda_1^{2b_{SO_1}} g^2 v^2\,,
$$
which implies that
$$
g = (-1)^{b_{SO_1}}x\,,
$$
and
$$
C = B_1(0).
$$
If we set $C = g B_1(0)$ in (\ref{CCCC}), then $g=1$
and $x = (-1)^{b_{SO_1}}$.

In any case, we have $g^2(\tau_2)=1$, which,
by the Jacobi identity, is equivalent to
${\vartheta_3^4(\tau)}{\vartheta_4^4(\tau)}
= 0$.
As we saw in the previous section,
$\tau_2=0$ is a solution of the above equation,
which provides a consistency check on procedure.

In summary, we have seen that we can lift the middle
1/2NS5-branes out of the configuration provided
that $b_{Sp_2}=2k_2+2-2k_1=0$,
which is equivalent to
$$
k_2 = k_1 - 1\,.
$$
If we look up the $O4^-$-plane in table (\ref{gensetup}),
we see that this corresponds to $N'=N-1$.
Therefore, this confirms the result listed in
table (\ref{result1}) that a $D4$-brane is annihilated when a physical
NS5-brane splits on an $O4^-$-plane.


\subsection{The splitting on $\widetilde{O4^\pm}$-planes}
As we have argued in section \ref{result},
a physical NS-brane splits on an $\widetilde{O4^+}$-plane
with neither creation nor annihilation of $D4$-branes and
likewise it can split on an $\widetilde{O4^-}$-plane
with the annihilation of a $D4$-brane.
In order to obtain this result, it was necessary to
take into account the energy cost due to the bending of the
1/2NS-branes.

In particular, the minimum energy
configuration after the splitting corresponds to the 1/2NS-branes
intersecting each other at some distance from the $O4$-plane.
Note that this system does not correspond to an asymptotically
free theory and therefore we do not expect to see any splitting
in the range of validity of the Seiberg-Witten description.
This is what we are about to show. In other words, we will show that
another choice in table \ref{resambiguous}, which correspond to
asymptotically free theory (i.e., $1$ for $\widetilde{O4^+}$
and $0$ for $\widetilde{O4^-}$), can not be unsplitted and lifed away
from the orientifold planes.

For simplicity, we present only the case of
$SO(2k_1+1)\times Sp'(k_2) \times SO(2k_3+1)$.
The beta-function coefficients are
\be
\label{threeb_1}
b_{SO_1}=(2k_1+1)-2-k_2,~~~
b_{Sp_2}=2k_2+2-(k_1+k_3+1),~~~~
b_{SO_3}=(2k_3+1)-2-k_2,
\ee
Both $SO(2k_1+1)$
and $SO(2k_3+1)$ have $2k_2$ half-hypers or $k_2$ hypers. On the
other hand, $Sp'(k_2)$ has $(2k_1+1)+(2k_3+1)$ half-hypers,
for a total $k_1+k_3+1$ hypers.
The curve for this gauge theory is given by Eq.(3.42) in \cite{Land}
$$
vt^4+B_1(v^2) t^3+ v B_2(v^2) t^2+ B_3(v^2) t+v=0
$$
Assigning the value $2k_1-1$ to the dimension of $t$,
we find the curve to be
$$
vt^4+B_1(v^2) t^3+ c_1 \Lambda_1^{2b_{SO_1}} v B_2(v^2) t^2 +
$$
\be
\label{odd_odd}
c_2 \Lambda_1^{4b_{SO_1}} \Lambda_2^{2b_{Sp_2}}
B_3(v^2) t+
c_3 \Lambda_1^{6b_{SO_1}} \Lambda_2^{4b_{Sp_2}}
\Lambda_3^{2b_{SO_3}} v=0\,,
\ee
and taking the various decoupling limits, we determine
the coefficients to be
\be
\label{cvalue_1}
c_1=16(-1)^{b_{SO_1}+1},~~~c_2= 16^3  (-1)^{b_{Sp_2}}\,.~~~
c_3=16^6
\ee

\par
Unlike the previous subsections, the curve (\ref{odd_odd}) is invariant
under $v\rightarrow -v,t\rightarrow -t$. Equivalently, the orientifold
action is $X^{4,5}\rightarrow -X^{4,5}$, $X^{6}\rightarrow X^{6}$
and $X^{10}\rightarrow X^{10}+\pi R$. The symmetrized
form of (\ref{odd_odd}) is given by rescaling
$t\rightarrow At\equiv 2^6 i\Lambda^{2b_{SO}}\Lambda_2^{b_{Sp_2}}$
under the contraints $k_1=k_3$ and
$\Lambda_1=\Lambda_3=\Lambda$. We find
\be
\label{sym_odd_odd}
{\cal P}(v^2,t) = v A^4 t^4+B_1(v^2) A^3 t^3 -
c_1 \Lambda_1^{2b_{SO_1}} v B_2(v^2)A^2 t^2+
B_3(v^2) A^3 t +v A^4=0\,.
\ee
the curve ${\cal P}(v^2,t)$ has
a $t \to 1/t$ symmetry
$$
t^4 {\cal P}(v^2,1/t) = {\cal P}(v^2,t)\,.
$$
In the limit where the middle NS5-branes are lifted ${\cal P}(v^2,t)$,
reduces to
$$
{\cal P}(v^2,t) = {\cal R}(t){\cal Q}(v^2,t),
$$
where the curve ${\cal Q}(v^2,t)$ describes an $SO(2r+1)$
curve with no flavours.

We have that
$$
t^4 {\cal P}(v^2,1/t) = t^4 {\cal R}(1/t) {\cal Q}(v^2,1/t)
= {\cal P}(v^2,t) = {\cal R}(t){\cal Q}(v^2,t)\,
$$
is solved by ${\cal R}(t)=(t+x)^2$, where $x =\pm 1$.
Therefore, setting
$$
{\cal P}(v^2,t) = (t + x)^2
\left( 2^8 \Lambda_1^{2b_{SO_1}}\Lambda_2^{2b_{Sp_2}} v t^2 +
\alpha \tilde{B}(v^2) t
+ 2^8 \Lambda_1^{2b_{SO_1}}\Lambda_2^{2b_{Sp_2}} v \right)\,,
$$
implies the following identities
$$
- 4 i\Lambda_2^{b_{Sp_2}} B_1(v^2) =
\alpha \tilde{B}(v^2) + 2^9 x
\Lambda_1^{2b_{SO_1}}\Lambda_2^{2b_{Sp_2}} v,
$$
$$
(-1)^{b_{SO_1}+1} v B_2(v^2) =  2^9
\Lambda_1^{2b_{SO_1}}\Lambda_2^{2b_{Sp_2}} v
+ 2 x \alpha \tilde{B}(v^2)\,.
$$
In particular we get
$$
(-1)^{b_{SO_1}+1} v B_2(v^2) -
2^9 \Lambda_1^{2b_{SO_1}}\Lambda_2^{2b_{Sp_2}} v =
- 8 x i \Lambda_2^{b_{Sp_2}} B_1(v^2) -
2^{10} \Lambda_1^{2b_{SO_1}}\Lambda_2^{2b_{Sp_2}} v\,.
$$
We see that the above equation cannot be satisfied since the
two polynomials have different degrees. On the l.h.s we have
a polynomial of odd degree in $v$, whereas on the r.h.s. the degree
is even.
Therefore, we conclude that for $k_1-1/2
\leq k_2 \leq 2k_1-1$, it is not possible to lift
the middle 1/2NS5-branes.

As we remarked at the beginning,
this result tells us that another choice in table \ref{resambiguous}
is not correct and leaves us with the one given in table \ref{result1}.

\section{Conclusion}
In this paper, we have studied the general splitting process of
NS-branes and $D(p+2)$-branes on $Op$-planes with $p\leq 6$. Using
several arguments, we showed that for $D(p+2)$-brane splitting,
the rule is universal, while for NS-brane splitting, the rule is
not universal and depends on the dimension $p$. The results are
summarized in table \ref{result1} in the introduction.

\par
In a special case, namely for NS-branes on $O4$-planes, we can
study the process with the appropriate Seiberg-Witten curves. If
the splitting is allowed, the reverse process (unsplitting) must
also take place. This corresponds to a particular strong coupling
limit on the field theory side and was checked explicitly using
the curves.

\par
There are a lot of directions we can follow. First, it would still
be very nice if we could derive these results directly, for
example, from a world-sheet CFT calculation. Second, we could
generalize the above discussion to M-theory and consider, for
example, the splitting of M5-branes on OM2-planes. There are also
other orientifold-like planes, for example $ON5$-planes, which are
S-dual to $O5$-planes, and $OF1_{A,B}$-planes. Third, in
\cite{Kap,Han2} the mirror pairs of $Sp(k)$ groups are derived by
using the $O5^-$ and its S-duality counterparts $ON^-$ and $ON^0$
(the bound state of $ON^-$ and one physical NS-brane. See
\cite{9805019}, and some applications in \cite{9906031,9909125}).
It would be natural to use the $O5^+$-plane to study the mirror of
$SO(n)$ gauge groups. The problem with this approach is that we do
not know the action of S-duality on the $ON^+$-plane. Fourth, we
have predicted the existence of a 1/4$D5$-brane when NS-branes
split on $\widetilde{O5^\pm}$-planes and argued that it does not
cause any inconsistency. However, further study is still very
welcome to clarify this point. Finally, it would be very
interesting to find some applications of the above non-trivial
string dynamical process.

\*{\bf Acknowledgments}.
We are grateful to Yang Hui He and Asad Naqvi for taking part in
the project at an early stage.

\section{Appendix A: The generalized supersymmetric configuration}

We have discussed the supersymmetric configurations of
NS-branes and $D5$-branes in the case of $O3$-planes. Now we want to see if
these results can be generalized to other $Op$-planes. We will give two
different arguments.

\par
The first one relies on the linking number. Let us focus on the
supersymmetric configuration containing $O4$-planes first. In this case,
$O4^+,\widetilde{O4^+}$ have charge $1/2$, $O4^-$ has charge $-1/2$
and $\widetilde{O4^-}$ has charge $0$. Using the linking number
$$
L_{1/2NS}=\frac{1}{2}(R_{D6}-L_{D6})+(L_{D4}-R_{D4})
$$
as in \cite{Mir}
we get the following constraints on the number of
$Dp$-branes between a 1/2NS-brane and a $1/2D5$-brane
\begin{equation}
\label{O4susy}
\begin{array}{ll}
(O4^- ----- \widetilde{O4^+}) ~~  \& ~~ (\widetilde{O4^+} ----- O4^-)
& ~~~~~ N+\widetilde{N}=0  \\
(O4^+ ----- \widetilde{O4^-})~~  \& ~~ (\widetilde{O4^-} ----- O4^+)
&~~~~~ N+\widetilde{N}=1
\end{array}
\end{equation}
Let us explain the above notation. For example, by
$(O4^- ----- \widetilde{O4^+})$ we mean the
following two brane setups:
going from left to right we have either
$(O4^-, 1/2NS-brane, O4^+ + N~Dp-branes,
1/2D(p+2)-brane,\widetilde{O4^+})$ or
$(O4^-,1/2D(p+2)-brane, \widetilde{O4^-}
+ \widetilde{N}~Dp-branes,1/2NS-brane
,\widetilde{O4^+})$. These two are related by shifting
the 1/2NS-brane and the 1/2$D(p+2)$-brane.
The above brane setup is supersymmetric
if and only if $N \geq 0, \widetilde{N}
\geq 0$ and they satisfy the equation (\ref{O4susy}).
Note that the answer is same as that for the $O3$-planes.
In fact, it can be checked by an explicit
calculation that the
above conclusion is the same for all $p\leq 5$.

\par
It is possible to retrieve this general result
by a T-duality argument. Let us
consider $O4$-planes again.
Starting from the $O3$-planes we perform a T-duality
along the $X^3$ coordinate
(the world volume coordinates of the various branes are:
$O3$-plane $X^{126}$, NS-brane $X^{12345}$  and $D5$-brane $X^{12789}$).
We draw the brane setup in the $(X^6,X^3)$ plane (see the left-hand
side of
Figure \ref{fig:Tdual}). Notice that in this brane setup
the 1/2NS-brane intersects
two orientifold planes, whereas the $1/2D5$-brane intersects only one
orientifold plane. After T-duality along $X^3$,
two $O3^-$-planes combine to give one $O4^-$-plane \cite{Han1},
two $O3^+$-planes combine to give one $O4^+$-plane,
while one $O3^+$-plane and one $\widetilde{O3^+}$-plane
combine to give one $\widetilde{O4^+}$-plane. Therefore, we get the
right-hand side of Figure \ref{fig:Tdual}. The crucial thing is that
under T-duality, the number of allowed $Dp$-branes between the 1/2NS-brane
and the 1/2$D(p+2)$-brane is invariant. This explain why we have
the above general result.
\EPSFIGURE[ht]{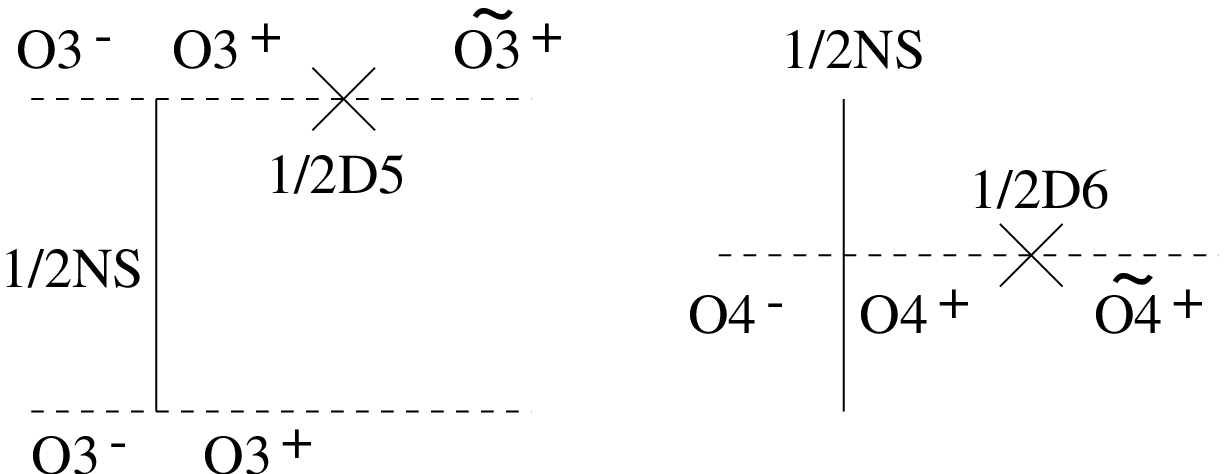,width=14cm}{The left-hand side is the correct
supersymmetric configuration of $O3$-planes. After T-duality, we get the
supersymmetric configuration for $O4$-planes at the right-hand side.
\label{fig:Tdual}}

\section{Appendix B: Ambiguity in the modular function}

The curves describing the Coulomb branch of
$SU(r+1)$, $Sp(r)$ and $SO(2r)$ theories in the conformal case
contain some modular functions of $\tau =
{\theta \over \pi} + i { 8 \pi \over g^2}$.
These expressions are determined recursively in the rank of
the group and by matching the Seiberg-Witten curve for
$SU(2)$ \cite{Arg2} \cite{Arg}.

We want to explain why there are actually two
physically equivalent expressions,
which differ by $\tau \to \tau + 1$.

The $SU(r+1)$ curve with $N_f = 2r+2$ reads
\cite{Arg}
$$
y^2 = \prod_{a=1}^{r+1} (x - \phi_a)^2 + 4h(h+1)
\prod_{j=1}^{2r+2} (x - m_j - 2h \mu)\,,
$$
where
\be
h(\tau) = { \vartheta_2^4 \over \vartheta_4^4 - \vartheta_2^4}\,.
\l{hey2}\ee

For $SO(2r+1)$ and $SO(2r)$, the modular function $f(\tau)$
in (\ref{tab_conf})
is expressed in terms of $h(\tau)$ \cite{Arg}.
This is achieved by matching the $SO(3)$ curve with
one vector hypermultiplet to the $SU(2)$ curve
with a massless adjoint hypermultiplet.

In particular
\be
f(\tau) = h(\tau) ( h(\tau)+1 )\,.
\l{heilla}\ee

In an earlier paper \cite{Arg2}, the authors analyzed
the curves for ${\cal N}=2$ $SU(r+1)$ gauge theories and
showed that
for the conformal curve you may choose two different
modular functions which are related to one another by
$\tau \to \tau + 1$. Both choices have the correct
weak-coupling asymptotic behaviour and match the Seiberg-Witten
curve for $SU(2)$. This amounts to a convential
choice of the origin of the $\theta$ angle.
This implies that, in (\ref{heilla}),
we can take either $h(\tau)$ or $h(\tau + 1)$.
Therefore, $f(\tau)$ or $f(\tau + 1)$ are physically
indistinguishable.

For $Sp(r)$, the unknown modular function $g(\tau)$
is recovered by matching to the Seiberg-Witten
solution for $SU(2) \cong Sp(1)$. The
map $\tau \to \tau + 1$ is a generator of
the duality group \cite{Arg}, under which
$g$ goes to $-g$. Then the
symplectic curve in (\ref{tab_conf}) is
invariant if the sign of a single mass is changed.

\bibliographystyle{JHEP}

\end{document}